\documentclass{bmcart}

\usepackage[utf8]{inputenc} 

\usepackage{graphicx}
\usepackage{amsmath}
\usepackage{amssymb}
\usepackage{mathrsfs}
\usepackage{stfloats}
\usepackage{dsfont}
\usepackage{setspace}
\usepackage{array}
\usepackage{bbm}
\usepackage{caption}

\newtheorem{theorem}{Theorem}

\newtheorem{proposition}{Proposition}
\newtheorem{lemma}{Lemma}
\newtheorem{remark}{Remark}

\newtheorem{event}{Event}

\startlocaldefs
\endlocaldefs

\begin{document}

\begin{frontmatter}

\begin{fmbox}


\title{Reflection Probability in Wireless Networks with Metasurface-Coated Environmental Objects: An Approach Based on Random Spatial Processes}

\author[
   addressref={aff1},
   corref={aff1},                       
   email={marcodi.renzo@l2s.centralesupelec.fr}
]{\inits{MDR}\fnm{Marco} \snm{Di Renzo}}
\author[
   addressref={aff1},                   
   email={jian.song@l2s.centralesupelec.fr}   
]{\inits{JS}\fnm{Jian} \snm{Song}}


\address[id=aff1]{%
  \orgname{Laboratoire des Signaux et Syst\`emes, CNRS, CentraleSupelec, Univ Paris-Sud, Universit\'e Paris-Saclay},
  \street{Plateau du Moulon},
  \postcode{91192},
  \city{Gif-sur-Yvette},
  \cny{France}                          
}



\end{fmbox}


\begin{abstractbox}

\begin{abstract} 
An emerging and promising vision of wireless networks consists of coating the environmental objects with reconfigurable metasurfaces that are capable of modifying the radio waves impinging upon them according to the generalized law of reflection. By relying on tools from point processes, stochastic geometry, and random spatial processes, we model the environmental objects with a modified random line process of fixed length, and with random orientations and locations. Based on the proposed modeling approach, we develop the first analytical framework that provides one with the probability that a randomly distributed object that is coated with a reconfigurable metasurface acts as a reflector for a given pair of transmitter and receiver. In contrast to the conventional network setup where the environmental objects are not coated with reconfigurable metasurfaces, we prove that the probability that the typical random object acts as a reflector is independent of the length of the object itself. The proposed analytical approach is validated against Monte Carlo simulations, and numerical illustrations are given and discussed.
\end{abstract}




\end{abstractbox}
%

\end{frontmatter}



%

\section{Methods/Experimental}
The methods used in the paper are based on the mathematical tools of random spatial processes and stochastic geometry. A new analytical framework for performance analysis is introduced. The theoretical framework is validated against Monte Carlo simulations.

\section{Introduction} \label{Introduction} 
Future wireless networks will be more than allowing people, mobile devices, and objects to communicate \cite{1}. Future wireless networks will be turned into a distributed intelligent wireless communications, sensing, and computing platform, which, besides communications, will be capable of sensing the environment to provide context-awareness capabilities, of locally storing and processing data to enable its time critical and energy efficient delivery, of accurately localizing people and objects in harsh propagation environments. Future wireless networks will have to fulfill the challenging requirement of interconnecting the physical and digital worlds in a seamless and sustainable manner \cite{2}, \cite{3}. \\

To fulfill these challenging requirements, it is apparent that it is not sufficient anymore to rely solely on wireless networks whose logical operation is software-controlled and optimized \cite{4}. The wireless environment itself needs to be turned into a software-reconfigurable entity \cite{5}, whose operation is optimized to enable uninterrupted connectivity. Future wireless networks need a smart radio environment, i.e., a wireless environment that is turned into a reconfigurable space that plays an active role in transferring and processing information. \\

Different solutions towards realizing this wireless future are currently emerging \cite{6}-\cite{14}. Among them, the use of reconfigurable metasurfaces constitutes a promising and enabling solution to fulfill the challenging requirements of future wireless networks \cite{15}. Metasurfaces are thin metamaterial layers that are capable of modifying the propagation of the radio waves in fully customizable ways \cite{16}, thus owing the potential of making the transfer and processing of information more reliable \cite{17}. Also, they constitute a suitable distributed platform to perform low-energy and low-complexity sensing \cite{18}, storage \cite{19}, and analog computing \cite{20}. For this reason, they are particularly useful for improving the performance of non-line-of-sight transmission, e.g., to appropriately customize the impact of multipath propagation. \\

In \cite{14}, in particular, the authors have put forth a network scenario where every environmental object is coated with reconfigurable metasurfaces, whose response to the radio waves is programmed in software by capitalizing on the enabling technology and hardware platform currently being developed in \cite{21}. Current research efforts towards realizing this vision are, however, limited to implement hardware testbeds, e.g., reflect-arrays and metasurfaces, and on realizing point-to-point experimental tests \cite{6}-\cite{14}. To the best of the authors knowledge, notably, there exists no analytical framework that investigates the performance of large-scale wireless networks in the presence of reconfigurable metasurfaces. In the present paper, motivated by these considerations, we develop the first analytical approach that allows one to study the probability that a random object coated with a reconfigurable metasurface acts as a reflector according to the generalized laws of reflection \cite{16}. To this end, we capitalize on the mathematical tool of random spatial processes \cite{Blockages}, \cite{Baccelli}. \\

Random spatial processes are considered to be the most suitable analytical tool to shed light on the ultimate performance limits of innovative technologies when applied in wireless networks, and to guide the design of optimal algorithms and protocols for attaining such ultimate limits. Several recent results on the application of random spatial processes in wireless networks can be found in \cite{83}-\cite{135}. Despite the many results available, however, fundamental issues remain open \cite{125}. In the current literature, in particular, the environmental objects are modeled as entities that can only attenuate the signals, by making the links either line-of-sight or non-line-of-sight, e.g., \cite{103}-\cite{105}. Modeling anything else is acknowledged to be difficult. Just in \cite{107}, the authors have recently investigated the impact of reflections, but only based on conventional Snell's laws. This work highlights the analytical complexity, the relevance, and the non-trivial performance trade-offs: The authors emphasize that the obtained trends highly depend on the fact that the total distance of the reflected paths is almost always two times larger than the distance of the direct paths. This occurs because the angles of incidence and reflection are the same based on Snell's law. In the presence of reconfigurable metasurfaces, on the other hand, the random objects can optimize the reflected signals in anomalous directions beyond Snell's law. The corresponding achievable performance and the associated optimal setups are unknown. \\

Motivated by these considerations, we develop an analytical framework that allows one to quantify the probability that a random object coated with reconfigurable metasurfaces acts as a reflector for a given pair of transmitter and receiver. Even though reconfigurable metasurfaces can be used to control and customize the refractions from environmental objects, in the present paper we focus our attention on controlling and customizing only the reflections of signals, since refractions may be subject to severe signal's attenuation. Our proposed approach, in particular, is based on modeling the environmental objects with a modified random line process of fixed length, and with random orientations and locations. In contrast to the conventional network setup where the environmental objects are not coated with reconfigurable metasurfaces, we prove that the probability that the typical random object acts as a reflector is independent of the length of the object itself. The proposed analytical approach is validated against Monte Carlo simulations, and numerical illustrations are given and discussed. In the present paper, we limit ourselves to analyze a 2D network scenarios, but our approach can be applied to 3D network topologies as well. This non-trivial generalization is postponed to a future research work. \\

The remainder of the present paper is organized as follows. In Section 3, the system model is introduced. In Section 4, the problem is formulated in mathematical terms. In Section 5, the analytical framework of the reflection probability is described. In Section 6, numerical results are illustrated, and the proposed approach is validated against Monte Carlo simulations. Finally, Section 7 concludes the paper. \\

\begin{table}[!t]
	\caption{Main symbols and functions used throughout the paper.}
	\begin{tabular}{|l||l|} \hline
	Symbol/Function & Definition \\ \hline \hline
	$\Pr \left\{ A \right\}$ & Probability of Event $A$ \\ \hline
$\Pr \left\{ {\overline A } \right\}$ & Probability of complement of Event $A$ \\ \hline
	$H\left(  \cdot  \right)$, $\bar H\left(  \cdot  \right)$ & Heaviside function, complementary Heaviside function \\ \hline
	$\left( {{x_{{\text{Tx}}}},{y_{{\text{Tx}}}}} \right)$ & Location of the transmitter \\ \hline
    $\left( {{x_{{\text{Rx}}}},{y_{{\text{Rx}}}}} \right)$ & Location of the receiver \\ \hline
    $\left( {{x_{{\text{object}}}},{y_{{\text{object}}}}} \right)$ & Location of the center of the typical object \\ \hline
	$\left( {{x_{{\text{end1}}}},{y_{{\text{end1}}}}} \right)$, $\left( {{x_{{\text{end2}}}},{y_{{\text{end2}}}}} \right)$ & Coordinates of the end points of the typical object \\ \hline
	$L$ & Length of the typical object \\ \hline
    ${R_{{\text{net}}}}$ & Radius of the network \\ \hline
	\end{tabular} \label{Table_Notation}
\end{table}
\textit{Notation}: The main symbols and functions used in this paper are reported in Table \ref{Table_Notation}.

\section{System Model} \label{SystemModel}
We consider a wireless network on a bi-dimensional plane, where the transmitters and receivers are distributed independently of each other. Without loss of generality, the location of the transmitter and receiver of interest, i.e., the probe transmitter and receiver, are denoted by $\left( {{x_{{\text{Tx}}}},{y_{{\text{Tx}}}}} \right)$ and $\left( {{x_{{\text{Rx}}}},{y_{{\text{Rx}}}}} \right)$, respectively.

\begin{figure}[!t]
	\setlength{\captionmargin}{10.0pt}
	\centering
	\includegraphics[width=1.00\columnwidth]{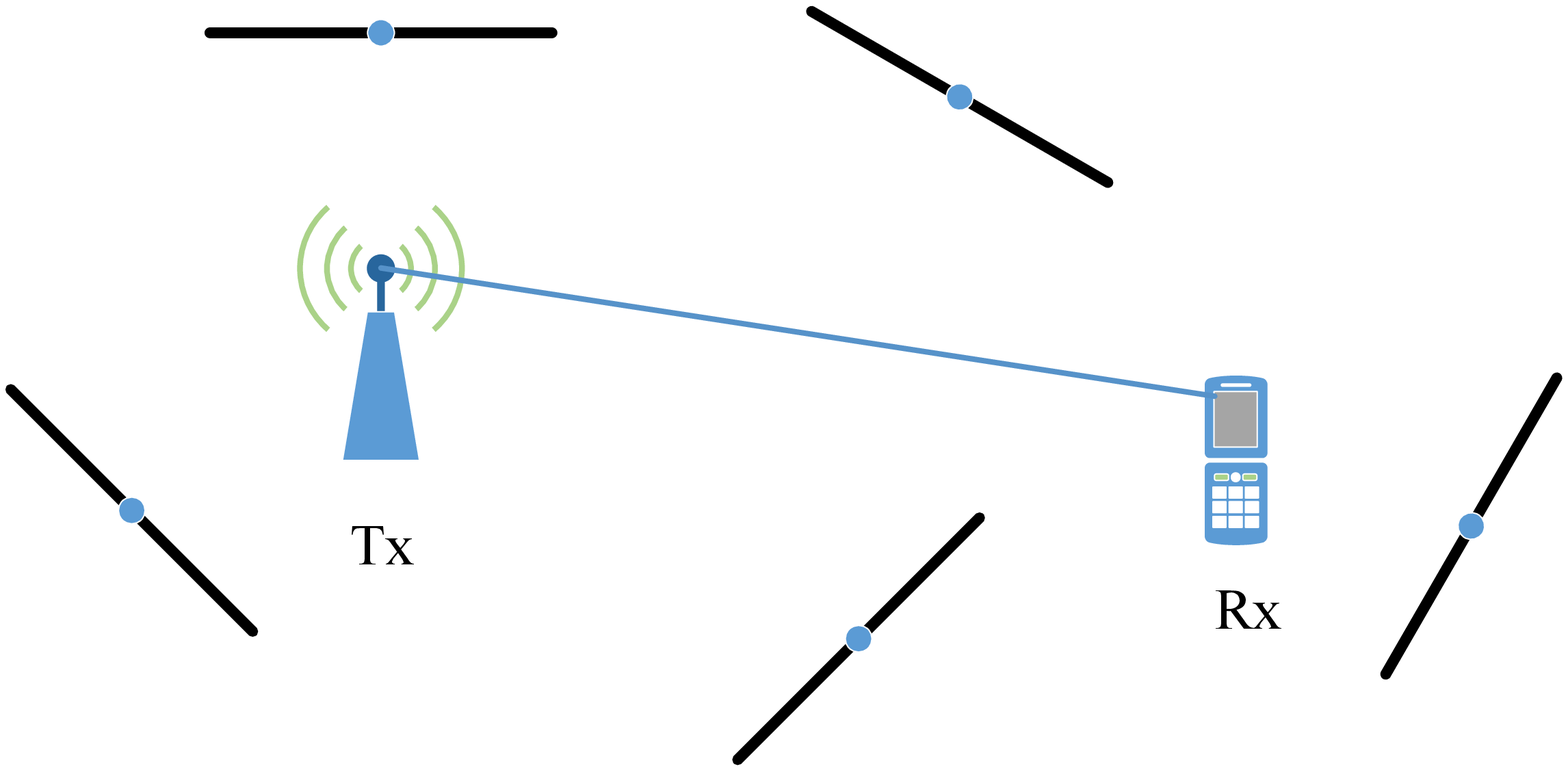}
	\caption{Probe transmitter (Tx) and receiver (Rx) in the presence of randomly distributed environmental objects.} \label{Fig_1}
\end{figure}
Besides the transmitter and receiver, we assume that environmental objects, e.g., buildings in a urban outdoor scenario, are randomly distributed in the same region. An example of the network model is depicted in Fig. \ref{Fig_1}. More precisely, the environmental objects are assumed to follow a Boolean model of line segments with the following properties \cite{Blockages}:
\begin{itemize}
	\item The center points of the objects form a homogeneous Poisson point process.
	\item The orientation of the objects are independent and identically distributed in $\left[ {0,2\pi } \right]$.
	\item The lengths of the objects are fixed and all equal to $L$.
    \item The random orientation and the center points of the objects are independent of each other. \\
\end{itemize}

We consider a generic environmental object, i.e., the typical object, and denote its center by $\left( {{x_{{\text{object}}}},{y_{{\text{object}}}}} \right)$, and the coordinates of its two end points by $\left( {{x_{{\text{end1}}}},{y_{{\text{end1}}}}} \right)$ and $\left( {{x_{{\text{end2}}}},{y_{{\text{end2}}}}} \right)$.

\section{Problem Formulation} \label{ProblemFormulation}
The objective of the present paper is to compute the probability that a randomly distributed object can act as a reflector for the pair of transmitter and receiver located in $\left( {{x_{{\text{Tx}}}},{y_{{\text{Tx}}}}} \right)$ and $\left( {{x_{{\text{Rx}}}},{y_{{\text{Rx}}}}} \right)$, respectively. We analyze two case studies: \\
\begin{itemize}
  \item \textbf{Scenario I}: The first scenario corresponds to the case study where the typical object is coated with a reconfigurable metasurface, which can optimize the angle of reflection regardless of the angle of incidence \cite{16}.
  \item \textbf{Scenario II}: The second scenario corresponds to the case study where the typical object is not coated with a reconfigurable metasurface. This is the state-of-the-art scenario, where the angle of reflection needs to be equal to the angle of incidence according to Snell's law of reflection \cite{16}. \\
\end{itemize}

\begin{figure}[!t]
	\setlength{\captionmargin}{10.0pt}
	\centering
	\includegraphics[width=1.00\columnwidth]{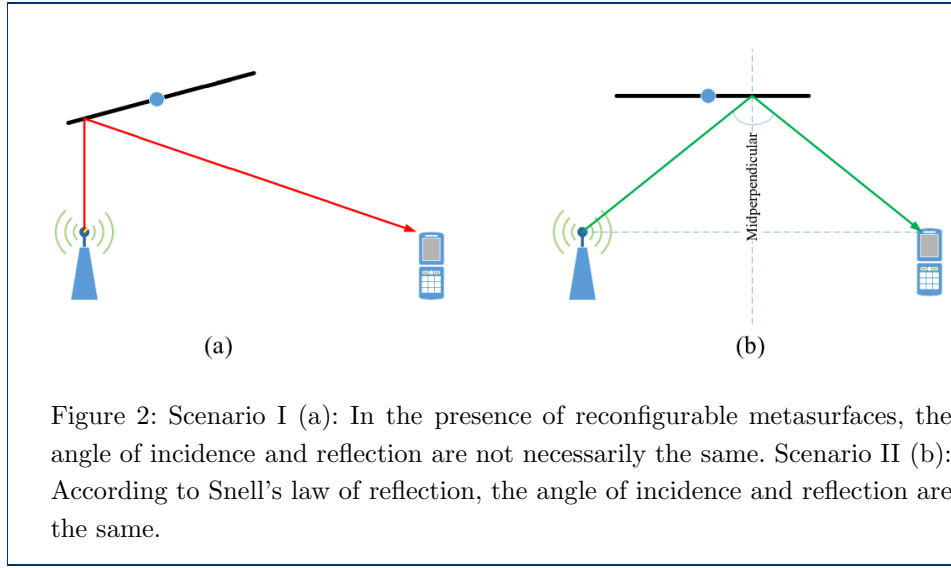}
	\caption{Scenario I (a): In the presence of reconfigurable metasurfaces, the angle of incidence and reflection are not necessarily the same. Scenario II (b): According to Snell's law of reflection, the angle of incidence and reflection are the same.} \label{Fig_2}
\end{figure}
The aim of the present paper is to develop a mathematical theory to compute the probability that the typical object can act as a reflector, i.e., the reflection probability, and to quantify the gain of adding reconfigurable metasurfaces in wireless networks. For analytical tractability, we assume that the reconfigurable metasurfaces are capable of producing any angle of reflection for any given location of transmitter and receiver, for any angle of incidence, and for any length. This yields the best-case performance bound compared with conventional Snell's law of reflection. The analysis, in addition, is conducted by relying on ray tracing arguments, in order to highlight the potential gains of using reconfigurable metasurfaces. The two case studies are sketched in Fig. \ref{Fig_2}. Generalizations of the proposed analytical framework are left to future research works.

\subsection{Scenario I: Reflections in the Presence of Reconfigurable Metasurfaces} \label{Scenario_I}
In the presence of reconfigurable metasurfaces, an arbitrary angle of reflection can be obtained for any angle of incidence. This implies that the typical object acts as a reflector for a transmitter and receiver if they are both located on the same side of the infinite line passing through the end points $\left( {{x_{{\text{end1}}}},{y_{{\text{end1}}}}} \right)$ and $\left( {{x_{{\text{end2}}}},{y_{{\text{end2}}}}} \right)$ of the typical object. \\

For ease of exposition, we introduce the following event. \\
\begin{event} \label{Event1}
The probe transmitter, Tx, and receiver, Rx, are located on the same side of the infinite line passing through the end points $\left( {{x_{{\rm{end1}}}},{y_{{\rm{end1}}}}} \right)$ and $\left( {{x_{{\rm{end2}}}},{y_{{\rm{end2}}}}} \right)$ of the typical object. \\
\end{event}

Therefore, the typical object acts as a reflector if Event \ref{Event1} holds true. Our objective is to formulate the probability of Event \ref{Event1}, i.e., to compute $\Pr \left\{ {{\text{Event 1}}} \right\}$. This latter probability can be formulated in two different but equivalent ways.

\begin{figure}[!t]
	\setlength{\captionmargin}{10.0pt}
	\centering
	\includegraphics[width=1.00\columnwidth]{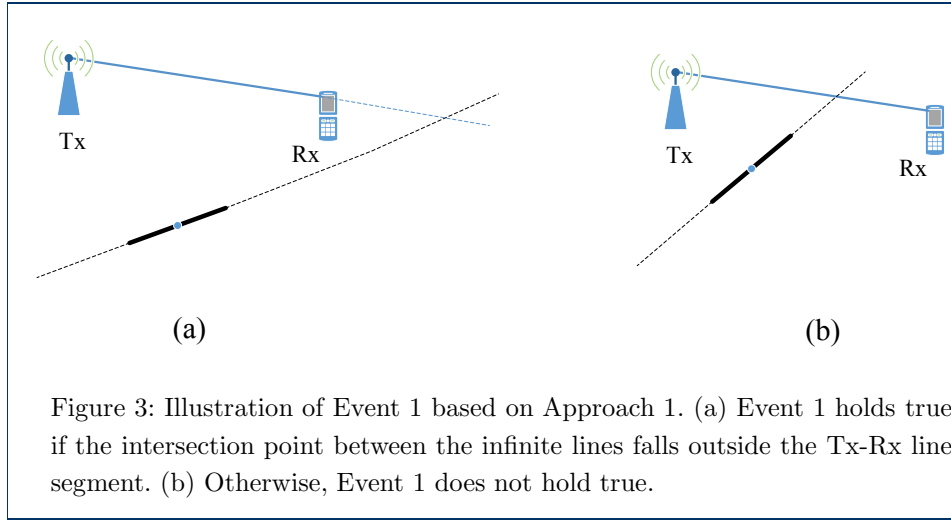}
	\caption{Illustration of Event 1 based on Approach 1. (a) Event 1 holds true if the intersection point between the infinite lines falls outside the Tx-Rx line segment. (b) Otherwise, Event 1 does not hold true.} \label{Fig_3}
\end{figure}
\subsubsection{Approach 1}
Let us consider the infinite line that connects the locations of transmitter and receiver, and the infinite line that connects the two end points of the typical object. Event 1 holds true if the intersection point, denoted by $\left( {{x^*},{y^*}} \right)$, of these two infinite lines falls outside the line segment that connects that transmitter and the receiver. An illustration is given in Fig. \ref{Fig_3}. \\

In mathematical terms, therefore, $\Pr \left\{ {{\text{Event 1}}} \right\}$ can be formulated as follows:
\begin{equation} \label{eq1}
\centering
\Pr \left\{ {{\text{Event 1}}} \right\} = 1 - \Pr \left\{ {\overline {{\text{Event 1}}} } \right\}
\end{equation}
where ${\overline {{\text{Event 1}}} }$ is the complement of ${{\text{Event 1}}}$, and $\Pr \left\{ {\overline {{\text{Event 1}}} } \right\}$ denotes the probability that the intersection point $\left( {{x^*},{y^*}} \right)$ is on the Tx-Rx line segment:
\begin{equation} \label{eq2}
\centering
\Pr \left\{ {\overline {{\text{Event 1}}} } \right\} = \Pr \left\{ \begin{gathered}
\min \left( {{x_{{\text{Tx}}}},{x_{{\text{Rx}}}}} \right) \leqslant {x^*} \leqslant \max \left( {{x_{{\text{Tx}}}},{x_{{\text{Rx}}}}} \right){\text{ }} \hfill \\
\cap {\text{ }}\min \left( {{y_{{\text{Tx}}}},{y_{{\text{Rx}}}}} \right) \leqslant {y^*} \leqslant \max \left( {{y_{{\text{Tx}}}},{y_{{\text{Rx}}}}} \right) \hfill \\
\end{gathered}  \right\}
\end{equation}

\begin{figure}[!t]
	\setlength{\captionmargin}{10.0pt}
	\centering
	\includegraphics[width=1.00\columnwidth]{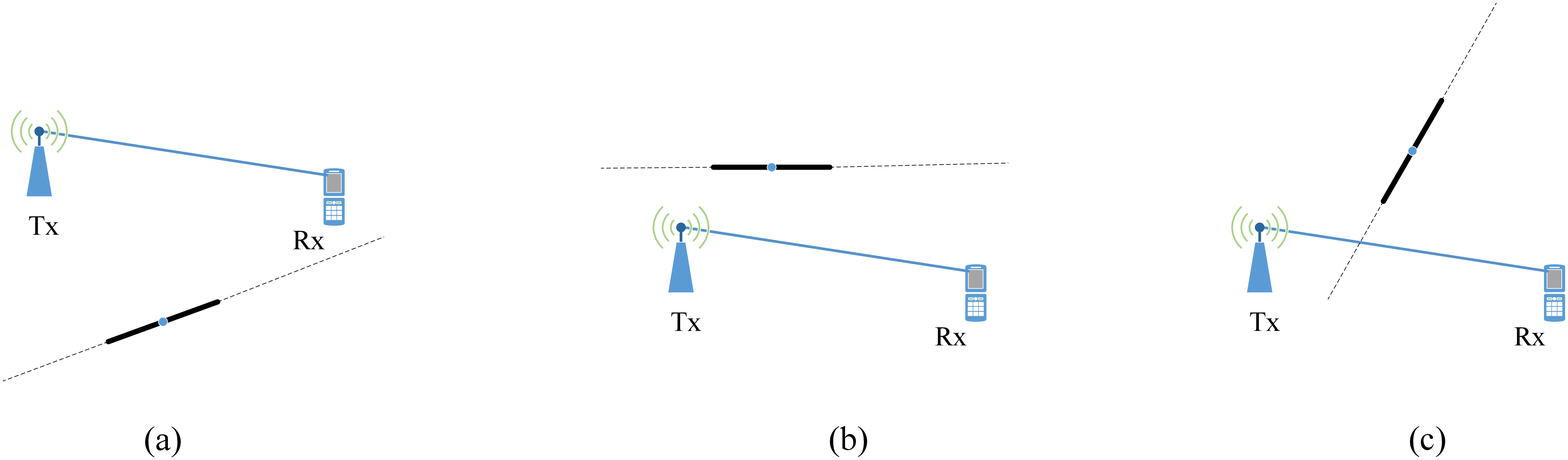}
	\caption{Illustration of Event 1 based on Approach 2. Event 1 holds true if both Tx and Rx are above (a) or below (b) the infinite line corresponding to the typical object. Event 1 does not hold true if Tx and Rx are not on the same side of the line (c).} \label{Fig_4}
\end{figure}
\subsubsection{Approach 2}
The probability that Event \ref{Event1} holds true can be formulated also in terms of the positions of the transmitter and receiver with respect to the infinite line that connects the end points of the typical object. In particular, Event \ref{Event1} holds true if both the transmitter and receiver are on the same side of the infinite line, i.e., either they are above or they are below the infinite line. This interpretation can be viewed as a problem of classifying points with respect to a line. This interpretation is depicted in Fig. \ref{Fig_4}. \\

In mathematical terms, therefore, $\Pr \left\{ {{\text{Event 1}}} \right\}$ can be formulated as follows:
\begin{equation} \label{eq3}
\centering
\Pr \left\{ {{\text{Event 1}}} \right\} = \Pr \left\{ \begin{gathered}
\left\{ {\left[ {{\text{Tx is above the line}}} \right]{\text{ }} \cap {\text{ }}\left[ {{\text{Rx is above the line}}} \right]} \right\} \hfill \\
\cup {\text{ }}\left\{ {\left[ {{\text{Tx is below the line}}} \right]{\text{ }} \cap {\text{ }}\left[ {{\text{Rx is below the line}}} \right]} \right\} \hfill \\
\end{gathered}  \right\}
\end{equation}

\subsection{Scenario II: Reflections in the Absence of Reconfigurable Metasurfaces} \label{Scenario_II}
In the absence of reconfigurable metasurfaces, the typical object acts as a reflector, for a given transmitter and receiver, only if the angles of reflection and incidence are the same. This is agreement with Snell's law of reflection, and imposes some geometric constraints among the locations of the typical object, the transmitter, and the receiver. In order to compute the corresponding probability of occurrence, we introduce the following event. \\

\begin{event} \label{Event2}
The mid-perpendicular of the line segment that connects the transmitter and receiver intersects the line segment that represents the typical object. \\
\end{event}

\begin{figure}[!t]
	\setlength{\captionmargin}{10.0pt}
	\centering
	\includegraphics[width=1.00\columnwidth]{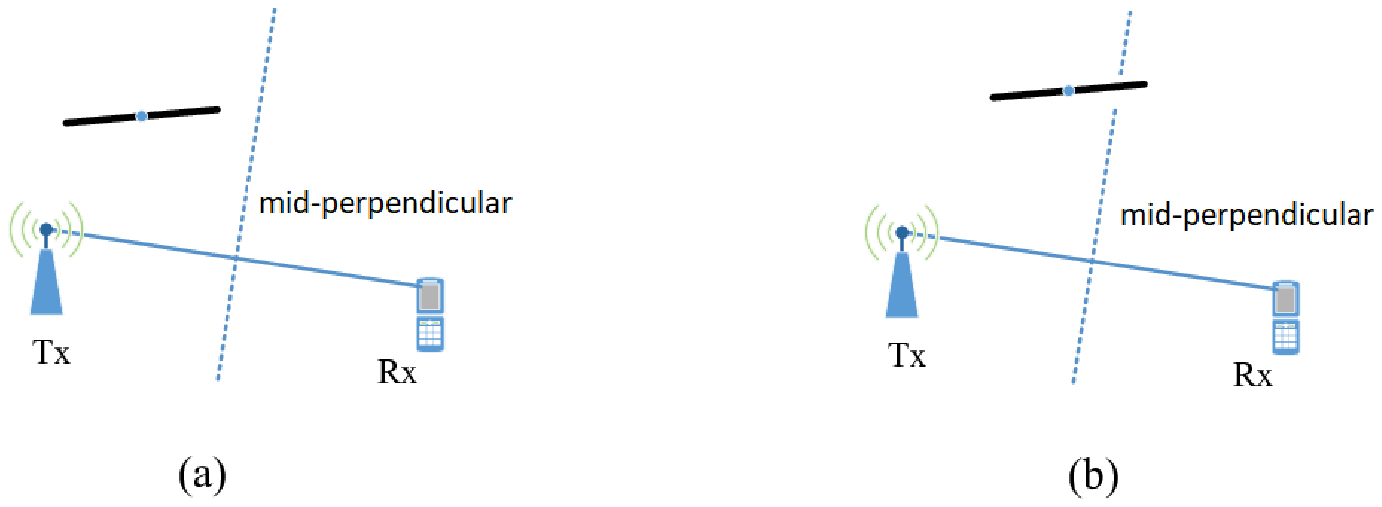}
	\caption{Illustration of Event 3. In (a), Event \ref{Event1} and Event \ref{Event2} hold true: The typical object acts as a reflector. In (b), Event \ref{Event1} holds true but Event \ref{Event2} does not hold true: The typical object cannot be a reflector.} \label{Fig_5}
\end{figure}
Let $\left( {{x_*},{y_*}} \right)$ denote the intersection between the mid-perpendicular of the line segment that connects the transmitter and receiver, and the line segment that represents the typical object. According to Snell's law of reflection, for some given locations of the transmitter and receiver, the typical object acts as a reflector if the mid-perpendicular of the line segment that connects the transmitter and receiver intersects the line segment that represents the typical object (i.e., Event \ref{Event2}), and, at the same time, the transmitted and receiver are located on the same side of the infinite line passing through the end points of the typical object (i.e., Event \ref{Event1}). \\

In mathematical terms, the probability of occurrence of Event \ref{Event2} can be formulated as follows:
\begin{equation} \label{eq4}
\centering
\Pr \left\{ {{\text{Event 2}}} \right\} = \Pr \left\{ \begin{gathered}
\min \left( {{x_{{\text{end1}}}},{x_{{\text{end2}}}}} \right) \leqslant {x_*} \leqslant \max \left( {{x_{{\text{end1}}}},{x_{{\text{end2}}}}} \right){\text{ }} \hfill \\
\cap {\text{ }}\min \left( {{y_{{\text{end1}}}},{y_{{\text{end2}}}}} \right) \leqslant {y_*} \leqslant \max \left( {{y_{{\text{end1}}}},{y_{{\text{end2}}}}} \right) \hfill \\
\end{gathered}  \right\}
\end{equation}

Based on Snell's law of reflection, therefore, the typical object acts a reflector if the following event holds true.\\

\begin{event} \label{Event3}
The transmitter and receiver are located on the same side of the infinite line passing through the end points of the typical object, and the mid-perpendicular of the line segment that connects the transmitter and receiver intersects the line segment that represents the typical object. \\
\end{event}

An illustration is given in Fig. \ref{Fig_5}. In mathematical terms, the probability of occurrence of Event \ref{Event1} can be formulated as follows: \\
\begin{equation} \label{eq5}
\Pr \left\{ {{\text{Event 3}}} \right\} = \Pr \left\{ {{\text{Event 1 }} \cap {\text{ Event 2}}} \right\}
\end{equation}
\section{Analytical Formulation of the Reflection Probability} \label{MathematicalFramework}
In this section, we introduce analytical expressions of the probability of occurrence of the three events introduced in the previous sections, and, therefore, characterize the probability that the typical object acts as a reflector in the presence and in the absence of reconfigurable metasurfaces. First, we begin with some preliminary results.
\subsection{Preliminary Results}

\begin{lemma} \label{lemma1}
Let $\left( {{x_{{\rm{Tx}}}},{y_{{\rm{Tx}}}}} \right)$ and $\left( {{x_{{\rm{Rx}}}},{y_{{\rm{Rx}}}}} \right)$ be the locations of the probe transmitter and receiver, respectively. The infinite line passing through them can be formulated as follows:
\begin{equation} \label{eq6}
\centering
y = mx + z
\end{equation}
where $m = \frac{{{y_{{\rm{Tx}}}} - {y_{{\rm{Rx}}}}}}{{{x_{{\rm{Tx}}}} - {x_{{\rm{Rx}}}}}}$, and $z = {y_{{\rm{Rx}}}} - m{x_{{\rm{Rx}}}}$. \\

\textit{Proof}: It follows by definition of line passing through two points. \hfill $\Box$ \\
\end{lemma}

\begin{figure}[t!]
	\setlength{\captionmargin}{10.0pt}
	\centering
	\includegraphics[width=0.70\columnwidth]{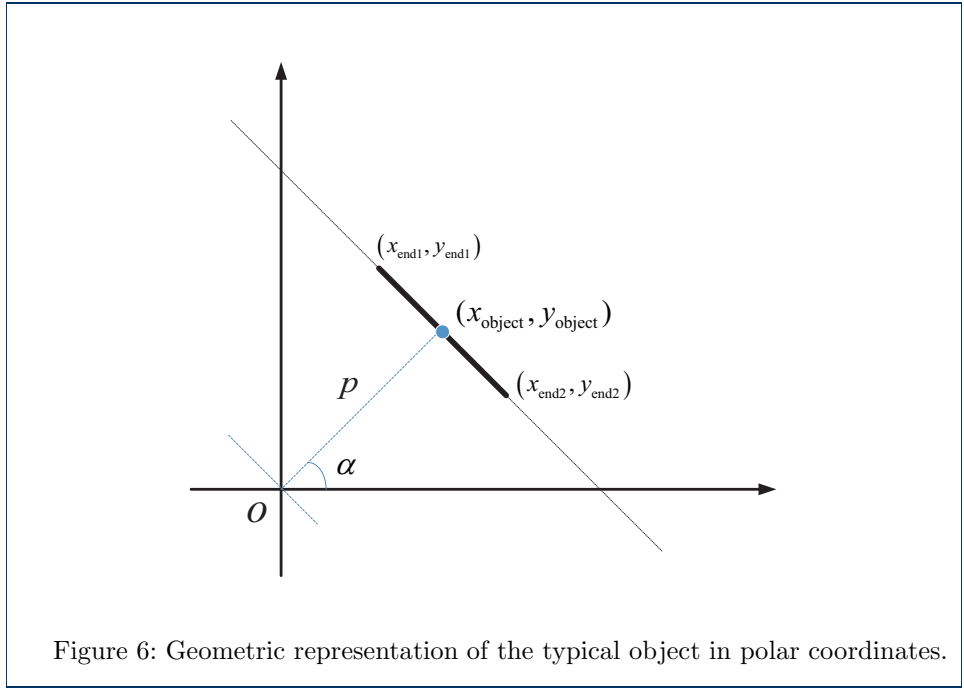}
	\caption{Geometric representation of the typical object in polar coordinates.} \label{Fig_6}
\end{figure}
\begin{lemma} \label{lemma2}
Let us consider the typical object of length $L$ depicted in Fig. \ref{Fig_6}. The distance between the center of the line segment and the origin is $p = {R_{{\rm{net}}}}\sqrt u $, where $u$ is a uniformly distributed random variable in $\left[ {0,1} \right]$, and ${R_{{\rm{net}}}}$ is largest size of the region of interest. Let $\alpha$ be the angle between the perpendicular line to the line segment, which passes through the center of the object, and the horizontal axis. The infinite line passing through the end points of the object can be formulated as follows:
\begin{equation} \label{eq7}
\centering
x\cos \alpha  + y\sin \alpha  = p
\end{equation}
where $\alpha \in \left[ {0,2\pi } \right]$. \\

In addition, the center of the line segment can be written as $\left( {{x_{{\rm{object}}}},{y_{{\rm{object}}}}} \right) = \left( {p\cos \alpha ,p\sin \alpha } \right)$, and its end points $\left( {{x_{{\rm{end1}}}},{y_{{\rm{end1}}}}} \right)$ and $\left( {{x_{{\rm{end2}}}},{y_{{\rm{end2}}}}} \right)$ can be formulated as follows:
\begin{equation} \label{eq8}
\centering
\begin{gathered}
{x_{{\rm{end1}}}} = {x_{{\rm{object}}}} - \frac{L}{2}\sin \alpha ,{\text{ }}{y_{{\rm{end1}}}} = {y_{{\rm{object}}}} + \frac{L}{2}\cos \alpha  \hfill \\
{x_{{\rm{end2}}}} = {x_{{\rm{object}}}} + \frac{L}{2}\sin \alpha ,{\text{ }}{y_{{\rm{end2}}}} = {y_{{\rm{object}}}} - \frac{L}{2}\cos \alpha  \hfill \\
\end{gathered}
\end{equation}

\textit{Proof}: The proof follows by noting that the centers of the line segments (the objects) are distributed according to a Poisson point process with random orientations, which implies $p = {R_{{\rm{net}}}}\sqrt u $ and $\alpha \in \left[ {0,2\pi } \right]$. The rest follows from geometric considerations. \hfill $\Box$ \\
\end{lemma}

\begin{lemma} \label{lemma3}
The mid-perpendicular of the infinite line in \eqref{eq6} is as follows:
\begin{equation} \label{eq9}
\centering
y = {m_p}x + {z_p}
\end{equation}
where ${m_p} =  - \frac{1}{m}$, and ${z_p} = \frac{1}{{2m}}\left( {{x_{{\rm{Tx}}}} + {x_{{\rm{Rx}}}}} \right) + \frac{1}{2}\left( {{y_{{\rm{Tx}}}} + {y_{{\rm{Rx}}}}} \right)$. \\

\textit{Proof}: See Appendix A. \hfill $\Box$ \\
\end{lemma}

\begin{lemma} \label{lemma4}
The intersection point between the infinite line that connects the transmitter and the receiver, and the infinite line that connects the end points of the line segment representing the typical object can be formulated as follows:
\begin{equation} \label{eq10}
\centering
\begin{gathered}
{x^*} = \frac{{p - z\sin \alpha }}{{m\sin \alpha  + \cos \alpha }} \hfill \\
{y^*} = m{x^*} + z \hfill \\
\end{gathered}
\end{equation}

The intersection point between the (infinite) mid-perpendicular line to the line segment that connects the transmitter and the receiver, and the infinite line that connects the end points of the line segment representing the typical object can be formulated as follows:
\begin{equation} \label{eq11}
\centering
\begin{gathered}
{x_*} = \frac{{p - {z_p}\sin \alpha }}{{{m_p}\sin \alpha  + \cos \alpha }} \hfill \\
{y_*} = {m_p}{x_*} + {z_p} \hfill \\
\end{gathered}
\end{equation}

\textit{Proof}: Equation \eqref{eq10} follows by solving the system of equations in \eqref{eq6} and \eqref{eq7}. Equation \eqref{eq11} follows by solving the system of equations in \eqref{eq7} and \eqref{eq9}. \hfill $\Box$ \\
\end{lemma}

\begin{lemma} \label{lemma5}
Let a generic infinite line formulated as: $ax + by + c = 0$. The following holds true: \\
\begin{itemize}
\item The point $\left( {{x_1},{y_1}} \right)$ is above the line if $a{x_1} + b{y_1} + c > 0$ and $b > 0$, or if $a{x_1} + b{y_1} + c < 0$ and $b < 0$.
\item The point $\left( {{x_1},{y_1}} \right)$ is below the line if $a{x_1} + b{y_1} + c < 0$ and $b > 0$, or if $a{x_1} + b{y_1} + c > 0$ and $b < 0$. \\
\end{itemize}

\textit{Proof}: See Appendix B. \hfill $\Box$
\end{lemma}

\subsection{Scenario I: Reflection Probability in the Presence of Reconfigurable Metasurfaces} \label{Frame_ScenarioI}
Theorem \ref{theorem1} and Theorem \ref{theorem2} provide one with analytical expressions of the probability that the typical object acts as a reflector if it is coated with reconfigurable metasurfaces. Theorem \ref{theorem1} is computed based on Approach 1, and Theorem \ref{theorem2} based on the Approach 2. \\

\begin{theorem} \label{theorem1}
Based on Approach 1, the probability of occurrence of Event \ref{Event1} is as follows:
\begin{equation} \label{eq12}
\centering
\begin{gathered}
\Pr \left\{ {{\rm{Event 1}}} \right\} = 1 - \Pr \left\{ {\overline {{\rm{Event 1}}} } \right\} \hfill \\
= 1 - \frac{1}{{2\pi }}\left\{ \begin{gathered}
\int_0^{{\delta _{\rm{1}}}} {{\theta _1}\left( {\alpha ,{x_{{\rm{Tx}}}},{x_{{\rm{Rx}}}},{y_{{\rm{Tx}}}},{y_{{\rm{Rx}}}}} \right)} d\alpha  + \int_{{\delta _{\rm{2}}}}^{2\pi } {{\theta _1}\left( {\alpha ,{x_{{\rm{Tx}}}},{x_{{\rm{Rx}}}},{y_{{\rm{Tx}}}},{y_{{\rm{Rx}}}}} \right)} d\alpha  \hfill \\
+ \int_0^{{\delta _{\rm{1}}}} {{\theta _2}\left( {\alpha ,{x_{{\rm{Tx}}}},{x_{{\rm{Rx}}}},{y_{{\rm{Tx}}}},{y_{{\rm{Rx}}}}} \right)} d\alpha  + \int_{{\delta _{\rm{2}}}}^{2\pi } {{\theta _2}\left( {\alpha ,{x_{{\rm{Tx}}}},{x_{{\rm{Rx}}}},{y_{{\rm{Tx}}}},{y_{{\rm{Rx}}}}} \right)} d\alpha  \hfill \\
+ \int_{{\delta _{\rm{1}}}}^{{\delta _{\rm{2}}}} {{\theta _3}\left( {\alpha ,{x_{{\rm{Tx}}}},{x_{{\rm{Rx}}}},{y_{{\rm{Tx}}}},{y_{{\rm{Rx}}}}} \right)} d\alpha  +  + \int_{{\delta _{\rm{1}}}}^{{\delta _{\rm{2}}}} {{\theta _4}\left( {\alpha ,{x_{{\rm{Tx}}}},{x_{{\rm{Rx}}}},{y_{{\rm{Tx}}}},{y_{{\rm{Rx}}}}} \right)} d\alpha  \hfill \\
\end{gathered}  \right\} \hfill \\
\end{gathered}
\end{equation}
where the integral limits are defined as ${\delta _{\rm{1}}} = 2{\tan ^{ - 1}}\left( {m + \sqrt {1 + {m^2}} } \right)$, and ${\delta _{\rm{2}}} = 2\pi  + 2{\tan ^{ - 1}}\left( {m - \sqrt {1 + {m^2}} } \right)$, and the auxiliary functions are given in Table \ref{Table_Functions1}. \\

\textit{Proof}: See Appendix C. \hfill $\Box$ \\
\end{theorem}

\begin{table}[!t] \footnotesize
\centering
\caption{Auxiliary functions used in Theorem \ref{theorem1}.} 
\newcommand{\tabincell}[2]{\begin{tabular}{@{}#1@{}}#2\end{tabular}}
\begin{tabular}{|l|} \hline
\hspace{5.00cm} Function definition \\ \hline \hline
$f\left( {\alpha ,\xi } \right) = \frac{1}{{{R_{{\rm{net}}}}}}\left( {\left[ {m\sin \alpha  + \cos \alpha } \right]\xi  + z\sin \alpha } \right)$
\\
$g\left( {\alpha ,\omega } \right) = \frac{1}{{{R_{{\rm{net}}}}}}\left( {\frac{{\left[ {m\sin \alpha  + \cos \alpha } \right]\left[ {\omega  - z} \right]}}{m} + z\sin \alpha } \right)$
\\
$\Theta \left( {\alpha \left| {\begin{array}{*{20}{c}}
		{{\mu _1}}&{{\mu _2}} \\
		{{\mu _3}}&{{\mu _4}}
		\end{array}} \right.} \right) = \left[ {{{\left( {\min \left\{ {{\mu _1},{\mu _2},1} \right\}} \right)}^2} - {{\left( {\max \left\{ {{\mu _3},{\mu _4},0} \right\}} \right)}^2}} \right]H\left( {\min \left\{ {{\mu _1},{\mu _2},1} \right\} - \max \left\{ {{\mu _3},{\mu _4},0} \right\}} \right)$
\\
${\theta _1}\left( {\alpha ,{x_{{\rm{Tx}}}},{x_{{\rm{Rx}}}},{y_{{\rm{Tx}}}},{y_{{\rm{Rx}}}}} \right) = \Theta \left( {\alpha \left| {\begin{array}{*{20}{c}}
		{f\left( {\alpha ,\max \left( {{x_{{\rm{Tx}}}},{x_{{\rm{Rx}}}}} \right)} \right)}&{g\left( {\alpha ,\max \left( {{y_{{\rm{Tx}}}},{y_{{\rm{Rx}}}}} \right)} \right)} \\
		{f\left( {\alpha ,\min \left( {{x_{{\rm{Tx}}}},{x_{{\rm{Rx}}}}} \right)} \right)}&{g\left( {\alpha ,\min \left( {{y_{{\rm{Tx}}}},{y_{{\rm{Rx}}}}} \right)} \right)}
		\end{array}} \right.} \right) \times H\left( m \right)$
\\
${\theta _2}\left( {\alpha ,{x_{{\rm{BS}}}},{x_{{\rm{MT}}}},{y_{{\rm{BS}}}},{y_{{\rm{MT}}}}} \right) = \Theta \left( {\alpha \left| {\begin{array}{*{20}{c}}
		{f\left( {\alpha ,\max \left( {{x_{{\rm{BS}}}},{x_{{\rm{MT}}}}} \right)} \right)}&{g\left( {\alpha ,\min \left( {{y_{{\rm{BS}}}},{y_{{\rm{MT}}}}} \right)} \right)} \\
		{f\left( {\alpha ,\min \left( {{x_{{\rm{BS}}}},{x_{{\rm{MT}}}}} \right)} \right)}&{g\left( {\alpha ,\max \left( {{y_{{\rm{BS}}}},{y_{{\rm{MT}}}}} \right)} \right)}
		\end{array}} \right.} \right) \times \bar H\left( m \right)$
\\
${\theta _3}\left( {\alpha ,{x_{{\rm{BS}}}},{x_{{\rm{MT}}}},{y_{{\rm{BS}}}},{y_{{\rm{MT}}}}} \right) = \Theta \left( {\alpha \left| {\begin{array}{*{20}{c}}
		{f\left( {\alpha ,\min \left( {{x_{{\rm{BS}}}},{x_{{\rm{MT}}}}} \right)} \right)}&{g\left( {\alpha ,\min \left( {{y_{{\rm{BS}}}},{y_{{\rm{MT}}}}} \right)} \right)} \\
		{f\left( {\alpha ,\max \left( {{x_{{\rm{BS}}}},{x_{{\rm{MT}}}}} \right)} \right)}&{g\left( {\alpha ,\max \left( {{y_{{\rm{BS}}}},{y_{{\rm{MT}}}}} \right)} \right)}
		\end{array}} \right.} \right) \times H\left( m \right)$
\\
${\theta _4}\left( {\alpha ,{x_{{\rm{BS}}}},{x_{{\rm{MT}}}},{y_{{\rm{BS}}}},{y_{{\rm{MT}}}}} \right) = \Theta \left( {\alpha \left| {\begin{array}{*{20}{c}}
		{f\left( {\alpha ,\min \left( {{x_{{\rm{BS}}}},{x_{{\rm{MT}}}}} \right)} \right)}&{g\left( {\alpha ,\max \left( {{y_{{\rm{BS}}}},{y_{{\rm{MT}}}}} \right)} \right)} \\
		{f\left( {\alpha ,\max \left( {{x_{{\rm{BS}}}},{x_{{\rm{MT}}}}} \right)} \right)}&{g\left( {\alpha ,\min \left( {{y_{{\rm{BS}}}},{y_{{\rm{MT}}}}} \right)} \right)}
		\end{array}} \right.} \right) \times \bar H\left( m \right)$
\\
\hline
		
\end{tabular} \label{Table_Functions1} 
\end{table}

\begin{theorem} \label{theorem2}
Based on Approach 2, the probability of occurrence of Event \ref{Event1} is as follows:
\begin{equation} \label{eq13}
\begin{gathered}
\Pr \left\{ {{\rm{Event 1}}} \right\} \hfill \\
= \frac{1}{{2\pi }}\int_0^{2\pi } {{\rho _1}\left( {\alpha ,{x_{{\rm{Tx}}}},{y_{{\rm{Tx}}}},{x_{{\rm{Rx}}}},{y_{{\rm{Rx}}}}} \right)} d\alpha  + \frac{1}{{2\pi }}\int_0^{2\pi } {{\rho _2}\left( {\alpha ,{x_{{\rm{Tx}}}},{y_{{\rm{Tx}}}},{x_{{\rm{Rx}}}},{y_{{\rm{Rx}}}}} \right)} d\alpha  \hfill \\
\end{gathered}
\end{equation}
where the integrand functions are defined as:
\begin{equation} \label{eq14}
\small
\begin{gathered}
{\rho _1}\left( {\alpha ,{x_{{\rm{Tx}}}},{y_{{\rm{Tx}}}},{x_{{\rm{Rx}}}},{y_{{\rm{Rx}}}}} \right) = {\left[ {\min \left\{ {\frac{{{x_{{\rm{Tx}}}}\cos \alpha  + {y_{{\rm{Tx}}}}\sin \alpha }}{{{R_{{\rm{net}}}}}},\frac{{{x_{{\rm{Rx}}}}\cos \alpha  + {y_{{\rm{Rx}}}}\sin \alpha }}{{{R_{{\rm{net}}}}}},1} \right\}} \right]^2} \hfill \\
\times H\left( {\min \left\{ {\frac{{{x_{{\rm{Tx}}}}\cos \alpha  + {y_{{\rm{Tx}}}}\sin \alpha }}{{{R_{{\rm{net}}}}}},\frac{{{x_{{\rm{Rx}}}}\cos \alpha  + {y_{{\rm{Rx}}}}\sin \alpha }}{{{R_{{\rm{net}}}}}},1} \right\}} \right) \hfill \\
\end{gathered}
\end{equation}
and
\begin{equation} \label{eq15}
\small
\begin{gathered}
{\rho _2}\left( {\alpha ,{x_{{\rm{Tx}}}},{y_{{\rm{Tx}}}},{x_{{\rm{Rx}}}},{y_{{\rm{Rx}}}}} \right) = \left[ {1 - {{\left( {\max \left\{ {\frac{{{x_{{\rm{Tx}}}}\cos \alpha  + {y_{{\rm{Tx}}}}\sin \alpha }}{{{R_{{\rm{net}}}}}},\frac{{{x_{{\rm{Rx}}}}\cos \alpha  + {y_{{\rm{Rx}}}}\sin \alpha }}{{{R_{{\rm{net}}}}}},0} \right\}} \right)}^2}} \right] \hfill \\
\times H\left( {1 - \max \left\{ {\frac{{{x_{{\rm{Tx}}}}\cos \alpha  + {y_{{\rm{Tx}}}}\sin \alpha }}{{{R_{{\rm{net}}}}}},\frac{{{x_{{\rm{Rx}}}}\cos \alpha  + {y_{{\rm{Rx}}}}\sin \alpha }}{{{R_{{\rm{net}}}}}},0} \right\}} \right) \hfill \\
\end{gathered}
\end{equation}

\textit{Proof}: See Appendix D. \hfill $\Box$ \\
\end{theorem}

\begin{remark} \label{remark5}
Theorem \ref{theorem1} and Theorem \ref{theorem2} are two analytical formulations of the same event. In the sequel, we show that they coincide. \hfill $\Box$ \\
\end{remark}

\subsection{Scenario II: Reflection Probability in the Absence of Reconfigurable Metasurfaces} \label{Frame_ScenarioII}
The probability of occurrence of Event \ref{Event3} is not easy to compute. The reason is that Event \ref{Event3} is formulated in terms of the intersection of Event \ref{Event1} and Event \ref{Event2}, which are not independent. In order to avoid the analytical complexity that originates from the correlation between Event \ref{Event1} and Event \ref{Event2}, we propose a upper-bound to compute the probability of occurrence of Event \ref{Event3}. Before stating the main result, we introduce the following proposition that provides one with the probability of occurrence of Event \ref{Event2}. \\

\begin{table}[!t] \footnotesize
\centering
\caption{Auxiliary functions used in Proposition \ref{proposition1}.}
\newcommand{\tabincell}[2]{\begin{tabular}{@{}#1@{}}#2\end{tabular}}
\begin{tabular}{|l|} \hline
\hspace{6.00cm} Function definition \\ \hline \hline
$F\left( {\alpha ,t} \right) = \frac{1}{{{R_{{\rm{net}}}}}}\left( {t + \frac{{{z_p}\sin \alpha }}{{{m_p}\sin \alpha  + \cos \alpha }}} \right){\left( {\frac{1}{{{m_p}\sin \alpha  + \cos \alpha }} - \cos \alpha } \right)^{ - 1}}$
\\
$G\left( {\alpha ,v} \right) = \frac{1}{{{R_{{\rm{net}}}}}}\left( {v + \frac{{{m_p}{z_p}\sin \alpha }}{{{m_p}\sin \alpha  + \cos \alpha }} - {z_p}} \right){\left( {\frac{{{m_p}}}{{{m_p}\sin \alpha  + \cos \alpha }} - \sin \alpha } \right)^{ - 1}}$
\\
$\Gamma _1^a\left( \alpha  \right) = \Theta \left( {\alpha \left| {\begin{array}{*{20}{c}}
		{F\left( {\alpha , - \frac{L}{2}\sin \alpha } \right)}&{G\left( {\alpha ,\frac{L}{2}\cos \alpha } \right)} \\
		{F\left( {\alpha ,\frac{L}{2}\sin \alpha } \right)}&{G\left( {\alpha , - \frac{L}{2}\cos \alpha } \right)}
		\end{array}} \right.} \right)H\left( {\frac{1}{{{m_p}\sin \alpha  + \cos \alpha }} - \cos \alpha } \right)H\left( {\frac{{{m_p}}}{{{m_p}\sin \alpha  + \cos \alpha }} - \sin \alpha } \right)$
\\
$\Gamma _1^b\left( \alpha  \right) = \Theta \left( {\alpha \left| {\begin{array}{*{20}{c}}
		{F\left( {\alpha , - \frac{L}{2}\sin \alpha } \right)}&{G\left( {\alpha , - \frac{L}{2}\cos \alpha } \right)} \\
		{F\left( {\alpha ,\frac{L}{2}\sin \alpha } \right)}&{G\left( {\alpha ,\frac{L}{2}\cos \alpha } \right)}
		\end{array}} \right.} \right)H\left( {\frac{1}{{{m_p}\sin \alpha  + \cos \alpha }} - \cos \alpha } \right)\bar H\left( {\frac{{{m_p}}}{{{m_p}\sin \alpha  + \cos \alpha }} - \sin \alpha } \right)$
\\
$\Gamma _1^c\left( \alpha  \right) = \Theta \left( {\alpha \left| {\begin{array}{*{20}{c}}
		{F\left( {\alpha ,\frac{L}{2}\sin \alpha } \right)}&{G\left( {\alpha ,\frac{L}{2}\cos \alpha } \right)} \\
		{F\left( {\alpha , - \frac{L}{2}\sin \alpha } \right)}&{G\left( {\alpha , - \frac{L}{2}\cos \alpha } \right)}
		\end{array}} \right.} \right)\bar H\left( {\frac{1}{{{m_p}\sin \alpha  + \cos \alpha }} - \cos \alpha } \right)H\left( {\frac{{{m_p}}}{{{m_p}\sin \alpha  + \cos \alpha }} - \sin \alpha } \right)$
\\
$\Gamma _1^d\left( \alpha  \right) = \Theta \left( {\alpha \left| {\begin{array}{*{20}{c}}
		{F\left( {\alpha ,\frac{L}{2}\sin \alpha } \right)}&{G\left( {\alpha , - \frac{L}{2}\cos \alpha } \right)} \\
		{F\left( {\alpha , - \frac{L}{2}\sin \alpha } \right)}&{G\left( {\alpha ,\frac{L}{2}\cos \alpha } \right)}
		\end{array}} \right.} \right)\bar H\left( {\frac{1}{{{m_p}\sin \alpha  + \cos \alpha }} - \cos \alpha } \right)\bar H\left( {\frac{{{m_p}}}{{{m_p}\sin \alpha  + \cos \alpha }} - \sin \alpha } \right)$
\\
$\Gamma _2^a\left( \alpha  \right) = \Theta \left( {\alpha \left| {\begin{array}{*{20}{c}}
		{F\left( {\alpha , - \frac{L}{2}\sin \alpha } \right)}&{G\left( {\alpha , - \frac{L}{2}\cos \alpha } \right)} \\
		{F\left( {\alpha ,\frac{L}{2}\sin \alpha } \right)}&{G\left( {\alpha ,\frac{L}{2}\cos \alpha } \right)}
		\end{array}} \right.} \right)H\left( {\frac{1}{{{m_p}\sin \alpha  + \cos \alpha }} - \cos \alpha } \right)H\left( {\frac{{{m_p}}}{{{m_p}\sin \alpha  + \cos \alpha }} - \sin \alpha } \right)$
\\
$\Gamma _2^b\left( \alpha  \right) = \Theta \left( {\alpha \left| {\begin{array}{*{20}{c}}
		{F\left( {\alpha , - \frac{L}{2}\sin \alpha } \right)}&{G\left( {\alpha ,\frac{L}{2}\cos \alpha } \right)} \\
		{F\left( {\alpha ,\frac{L}{2}\sin \alpha } \right)}&{G\left( {\alpha , - \frac{L}{2}\cos \alpha } \right)}
		\end{array}} \right.} \right)H\left( {\frac{1}{{{m_p}\sin \alpha  + \cos \alpha }} - \cos \alpha } \right)\bar H\left( {\frac{{{m_p}}}{{{m_p}\sin \alpha  + \cos \alpha }} - \sin \alpha } \right)$
\\
$\Gamma _2^c\left( \alpha  \right) = \Theta \left( {\alpha \left| {\begin{array}{*{20}{c}}
		{F\left( {\alpha ,\frac{L}{2}\sin \alpha } \right)}&{G\left( {\alpha , - \frac{L}{2}\cos \alpha } \right)} \\
		{F\left( {\alpha , - \frac{L}{2}\sin \alpha } \right)}&{G\left( {\alpha ,\frac{L}{2}\cos \alpha } \right)}
		\end{array}} \right.} \right)\bar H\left( {\frac{1}{{{m_p}\sin \alpha  + \cos \alpha }} - \cos \alpha } \right)H\left( {\frac{{{m_p}}}{{{m_p}\sin \alpha  + \cos \alpha }} - \sin \alpha } \right)$
\\
$\Gamma _2^d\left( \alpha  \right) = \Theta \left( {\alpha \left| {\begin{array}{*{20}{c}}
		{F\left( {\alpha ,\frac{L}{2}\sin \alpha } \right)}&{G\left( {\alpha ,\frac{L}{2}\cos \alpha } \right)} \\
		{F\left( {\alpha , - \frac{L}{2}\sin \alpha } \right)}&{G\left( {\alpha , - \frac{L}{2}\cos \alpha } \right)}
		\end{array}} \right.} \right)\bar H\left( {\frac{1}{{{m_p}\sin \alpha  + \cos \alpha }} - \cos \alpha } \right)\bar H\left( {\frac{{{m_p}}}{{{m_p}\sin \alpha  + \cos \alpha }} - \sin \alpha } \right)$
\\
$\Gamma _3^a\left( \alpha  \right) = \Theta \left( {\alpha \left| {\begin{array}{*{20}{c}}
		{F\left( {\alpha ,\frac{L}{2}\sin \alpha } \right)}&{G\left( {\alpha ,\frac{L}{2}\cos \alpha } \right)} \\
		{F\left( {\alpha , - \frac{L}{2}\sin \alpha } \right)}&{G\left( {\alpha , - \frac{L}{2}\cos \alpha } \right)}
		\end{array}} \right.} \right)H\left( {\frac{1}{{{m_p}\sin \alpha  + \cos \alpha }} - \cos \alpha } \right)H\left( {\frac{{{m_p}}}{{{m_p}\sin \alpha  + \cos \alpha }} - \sin \alpha } \right)$
\\
$\Gamma _3^b\left( \alpha  \right) = \Theta \left( {\alpha \left| {\begin{array}{*{20}{c}}
		{F\left( {\alpha ,\frac{L}{2}\sin \alpha } \right)}&{G\left( {\alpha , - \frac{L}{2}\cos \alpha } \right)} \\
		{F\left( {\alpha , - \frac{L}{2}\sin \alpha } \right)}&{G\left( {\alpha ,\frac{L}{2}\cos \alpha } \right)}
		\end{array}} \right.} \right)H\left( {\frac{1}{{{m_p}\sin \alpha  + \cos \alpha }} - \cos \alpha } \right)\bar H\left( {\frac{{{m_p}}}{{{m_p}\sin \alpha  + \cos \alpha }} - \sin \alpha } \right)$
\\
$\Gamma _3^c\left( \alpha  \right) = \Theta \left( {\alpha \left| {\begin{array}{*{20}{c}}
		{F\left( {\alpha , - \frac{L}{2}\sin \alpha } \right)}&{G\left( {\alpha ,\frac{L}{2}\cos \alpha } \right)} \\
		{F\left( {\alpha ,\frac{L}{2}\sin \alpha } \right)}&{G\left( {\alpha , - \frac{L}{2}\cos \alpha } \right)}
		\end{array}} \right.} \right)\bar H\left( {\frac{1}{{{m_p}\sin \alpha  + \cos \alpha }} - \cos \alpha } \right)H\left( {\frac{{{m_p}}}{{{m_p}\sin \alpha  + \cos \alpha }} - \sin \alpha } \right)$
\\
$\Gamma _3^d\left( \alpha  \right) = \Theta \left( {\alpha \left| {\begin{array}{*{20}{c}}
		{F\left( {\alpha , - \frac{L}{2}\sin \alpha } \right)}&{G\left( {\alpha , - \frac{L}{2}\cos \alpha } \right)} \\
		{F\left( {\alpha ,\frac{L}{2}\sin \alpha } \right)}&{G\left( {\alpha ,\frac{L}{2}\cos \alpha } \right)}
		\end{array}} \right.} \right)\bar H\left( {\frac{1}{{{m_p}\sin \alpha  + \cos \alpha }} - \cos \alpha } \right)\bar H\left( {\frac{{{m_p}}}{{{m_p}\sin \alpha  + \cos \alpha }} - \sin \alpha } \right)$
\\
$\Gamma _4^a\left( \alpha  \right) = \Theta \left( {\alpha \left| {\begin{array}{*{20}{c}}
		{F\left( {\alpha ,\frac{L}{2}\sin \alpha } \right)}&{G\left( {\alpha , - \frac{L}{2}\cos \alpha } \right)} \\
		{F\left( {\alpha , - \frac{L}{2}\sin \alpha } \right)}&{G\left( {\alpha ,\frac{L}{2}\cos \alpha } \right)}
		\end{array}} \right.} \right)H\left( {\frac{1}{{{m_p}\sin \alpha  + \cos \alpha }} - \cos \alpha } \right)H\left( {\frac{{{m_p}}}{{{m_p}\sin \alpha  + \cos \alpha }} - \sin \alpha } \right)$
\\
$\Gamma _4^b\left( \alpha  \right) = \Theta \left( {\alpha \left| {\begin{array}{*{20}{c}}
		{F\left( {\alpha ,\frac{L}{2}\sin \alpha } \right)}&{G\left( {\alpha ,\frac{L}{2}\cos \alpha } \right)} \\
		{F\left( {\alpha , - \frac{L}{2}\sin \alpha } \right)}&{G\left( {\alpha , - \frac{L}{2}\cos \alpha } \right)}
		\end{array}} \right.} \right)H\left( {\frac{1}{{{m_p}\sin \alpha  + \cos \alpha }} - \cos \alpha } \right)\bar H\left( {\frac{{{m_p}}}{{{m_p}\sin \alpha  + \cos \alpha }} - \sin \alpha } \right)$
\\
$\Gamma _4^c\left( \alpha  \right) = \Theta \left( {\alpha \left| {\begin{array}{*{20}{c}}
		{F\left( {\alpha , - \frac{L}{2}\sin \alpha } \right)}&{G\left( {\alpha , - \frac{L}{2}\cos \alpha } \right)} \\
		{F\left( {\alpha ,\frac{L}{2}\sin \alpha } \right)}&{G\left( {\alpha ,\frac{L}{2}\cos \alpha } \right)}
		\end{array}} \right.} \right)\bar H\left( {\frac{1}{{{m_p}\sin \alpha  + \cos \alpha }} - \cos \alpha } \right)H\left( {\frac{{{m_p}}}{{{m_p}\sin \alpha  + \cos \alpha }} - \sin \alpha } \right)$
\\
$\Gamma _4^d\left( \alpha  \right) = \Theta \left( {\alpha \left| {\begin{array}{*{20}{c}}
		{F\left( {\alpha , - \frac{L}{2}\sin \alpha } \right)}&{G\left( {\alpha ,\frac{L}{2}\cos \alpha } \right)} \\
		{F\left( {\alpha ,\frac{L}{2}\sin \alpha } \right)}&{G\left( {\alpha , - \frac{L}{2}\cos \alpha } \right)}
		\end{array}} \right.} \right)\bar H\left( {\frac{1}{{{m_p}\sin \alpha  + \cos \alpha }} - \cos \alpha } \right)\bar H\left( {\frac{{{m_p}}}{{{m_p}\sin \alpha  + \cos \alpha }} - \sin \alpha } \right)$
\\
\hline

\end{tabular}
\label{Table_Functions2}
\end{table}
\begin{proposition} \label{proposition1}
The probability of occurrence of Event \ref{Event2} can be formulated as follows.
\begin{equation} \label{eq17}
\centering
\begin{gathered}
\Pr \left\{ {{\rm{Event 2}}} \right\} = \Pr \left\{ \begin{gathered}
\min \left( {{x_{{\rm{end1}}}},{x_{{\rm{end2}}}}} \right) \leqslant {x_*} \leqslant \max \left( {{x_{{\rm{end1}}}},{x_{{\rm{end2}}}}} \right){\rm{ }} \hfill \\
\cap {\rm{ }}\min \left( {{y_{{\rm{end1}}}},{y_{{\rm{end2}}}}} \right) \leqslant {y_*} \leqslant \max \left( {{y_{{\rm{end1}}}},{y_{{\rm{end2}}}}} \right) \hfill \\
\end{gathered}  \right\} \hfill \\
= \frac{1}{{2\pi }}\left\{ {\int_{\frac{{3\pi }}{2}}^{2\pi } {{\Gamma _1}\left( \alpha  \right)} d\alpha  + \int_\pi ^{\frac{{3\pi }}{2}} {{\Gamma _2}\left( \alpha  \right)} d\alpha  + \int_0^{\frac{\pi }{2}} {{\Gamma _3}\left( \alpha  \right)} d\alpha  + \int_{\frac{\pi }{2}}^\pi  {{\Gamma _4}\left( \alpha  \right)} d\alpha } \right\} \hfill \\
\end{gathered}
\end{equation}
where ${\Gamma _1}\left( \alpha  \right) = \Gamma _1^a\left( \alpha  \right) + \Gamma _1^b\left( \alpha  \right) + \Gamma _1^c\left( \alpha  \right) + \Gamma _1^d\left( \alpha  \right)$, ${\Gamma _2}\left( \alpha  \right) = \Gamma _2^a\left( \alpha  \right) + \Gamma _2^b\left( \alpha  \right) + \Gamma _2^c\left( \alpha  \right) + \Gamma _2^d\left( \alpha  \right)$, ${\Gamma _3}\left( \alpha  \right) = \Gamma _3^a\left( \alpha  \right) + \Gamma _3^b\left( \alpha  \right) + \Gamma _3^c\left( \alpha  \right) + \Gamma _3^d\left( \alpha  \right)$, and ${\Gamma _4}\left( \alpha  \right) = \Gamma _4^a\left( \alpha  \right) + \Gamma _4^b\left( \alpha  \right) + \Gamma _4^c\left( \alpha  \right) + \Gamma _4^d\left( \alpha  \right)$, which are all defined in Table \ref{Table_Functions2}. \\

\emph{Proof}: See Appendix E. \hfill $\Box$ \\
\end{proposition}
\begin{theorem} \label{theorem3}
The probability of occurrence of Event \ref{Event3} is upper-bounded as follows:
\begin{equation} \label{eq16}
\centering
\Pr \left\{ {{\rm{Event 3}}} \right\} \leqslant \min \left\{ {\Pr \left\{ {{\rm{Event 1}}} \right\},\Pr \left\{ {{\rm{Event 2}}} \right\}} \right\}
\end{equation}
where ${\Pr \left\{ {{\rm{Event 1}}} \right\}}$ is formulated in Theorem \ref{theorem1} or Theorem \ref{theorem2}, and ${\Pr \left\{ {{\rm{Event 2}}} \right\}}$ is given in Proposition \ref{proposition1}. \\

\textit{Proof}: The proof follows by applying the Frechet inequality \cite{MDR_Frechet}. \hfill $\Box$ \\
\end{theorem}

\begin{remark} \label{remarkMDR}
By comparing Theorem \ref{theorem1} and Theorem \ref{theorem2} against Theorem \ref{theorem3}, we observe that the probability of being a reflector highly depends on the length of the typical object if is it not coated with a reconfigurable metasurfaces, while it is independent of it if it is coated with a reconfigurable metasurface. This is a major benefit of using reconfigurable metasurfaces in wireless networks. This outcome is determined by the assumption that the metasurfaces can modify the angle of reflection regardless of their length. The analysis of the impact of the constraints imposed by the size of the metasurface on its capability of obtaining a given set of angles of reflection as a function of the angle of incidence is an open but very important research issue, which is left to future research. \hfill $\Box$ \\
\end{remark}
\section{Numerical Results and Discussion: Validation Against Monte Carlo Simulations} \label{NumericalResults} 

\begin{figure}[!t]
	\setlength{\captionmargin}{10.0pt}
	\centering
	\includegraphics[width=1.00\columnwidth]{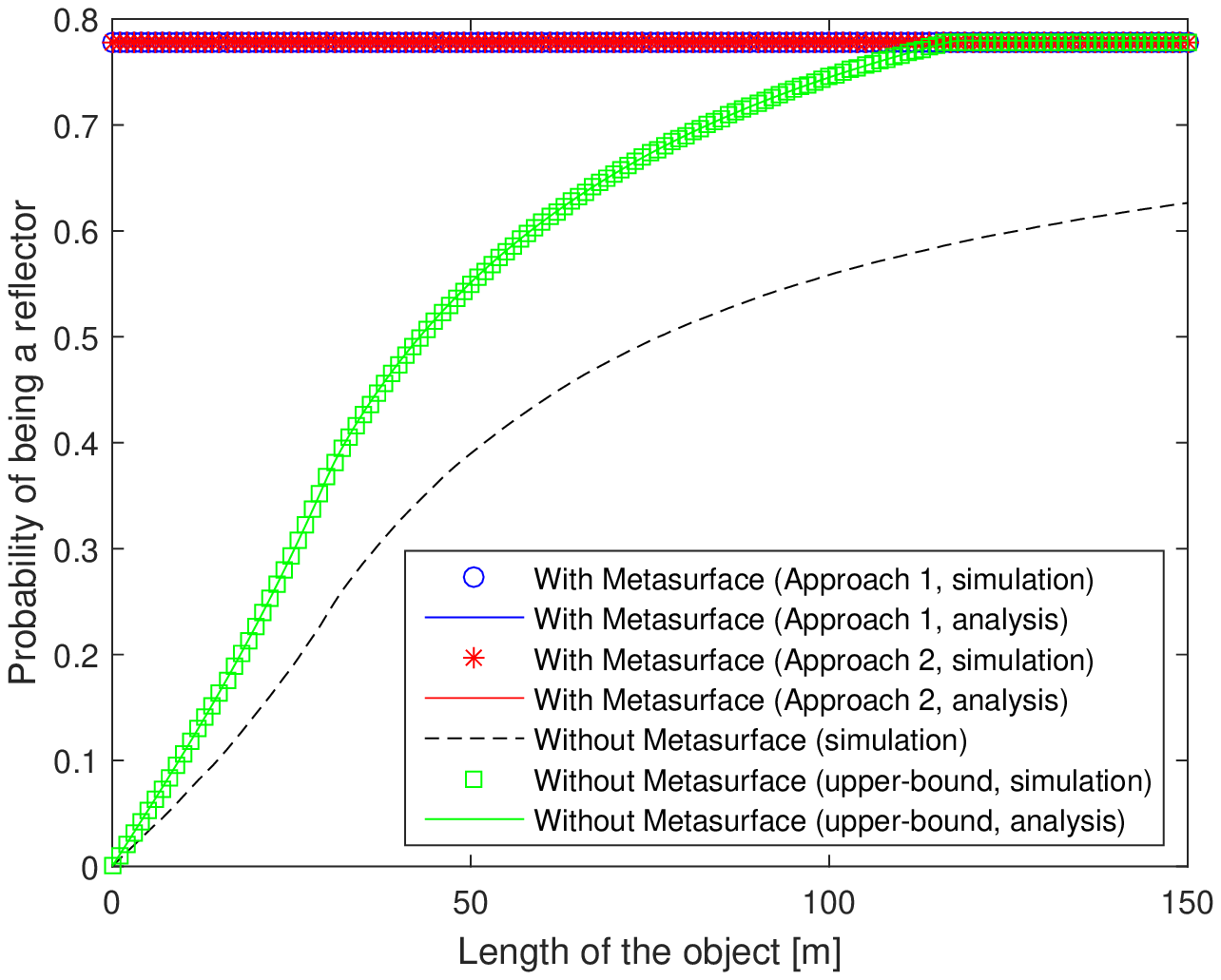}
	\caption{Probability of being a reflector versus the length of the object. Setup: ${R_{{\rm{net}}}} = 30{\rm{ m}}$, location of the transmitter $\left( {0,3} \right)$, location of the receiver $\left( {20,20} \right)$.} \label{Fig_8}
\end{figure}

\begin{figure}[!t]
	\setlength{\captionmargin}{10.0pt}
	\centering
	\includegraphics[width=1.00\columnwidth]{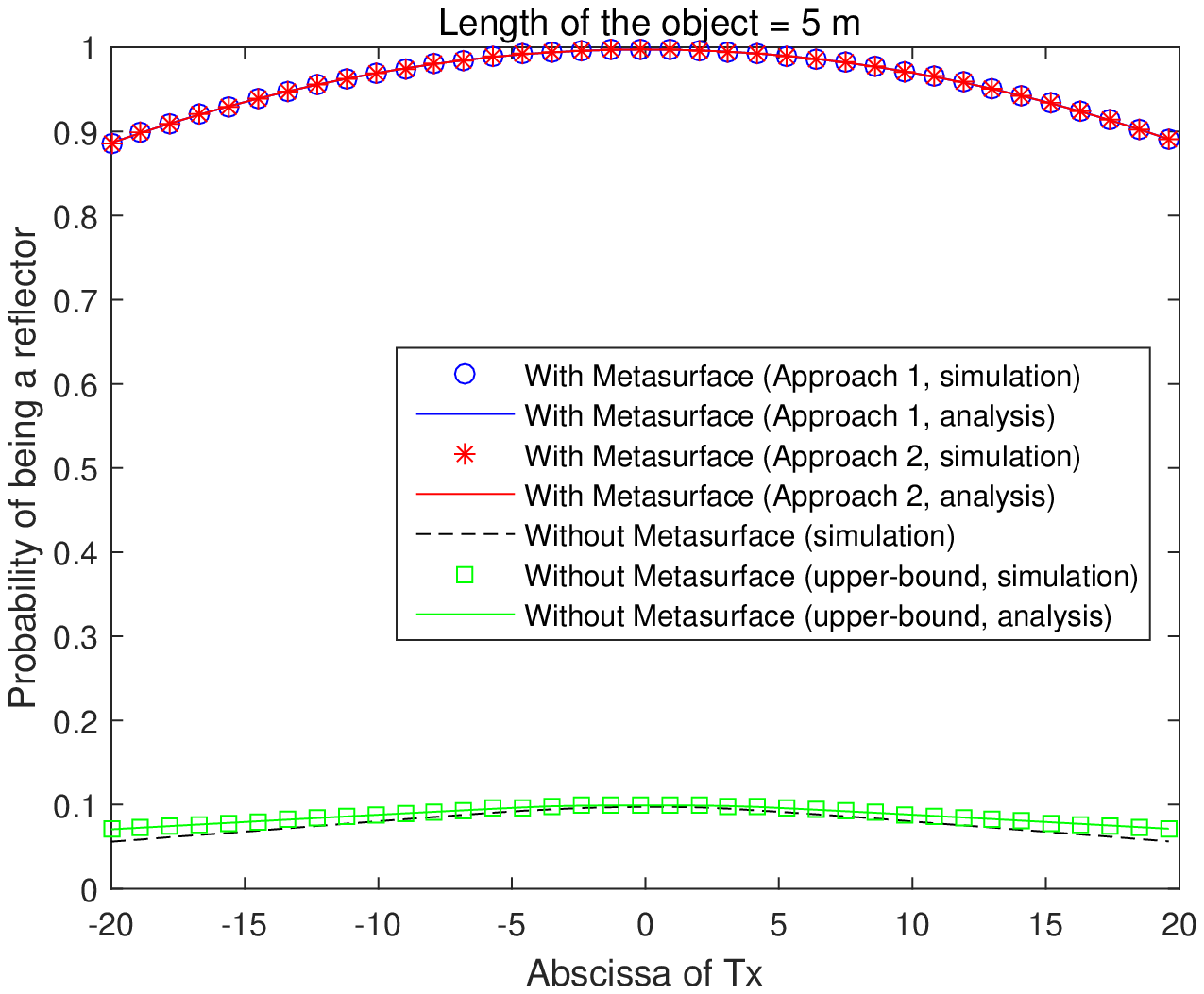}
	\caption{Probability of being a reflector versus the horizontal location, ${x_{{\rm{Tx}}}}$, of the transmitter. Setup: ${R_{{\rm{net}}}} = 30{\rm{ m}}$, vertical location of the transmitter ${y_{{\rm{Tx}}}} = 3$, location of the receiver $\left( {0,0} \right)$, length of the object $L = 5{\rm{ m}}$.} \label{Fig_9}
\end{figure}

\begin{figure}[!h]
	\setlength{\captionmargin}{10.0pt}
	\centering
	\includegraphics[width=1.00\columnwidth]{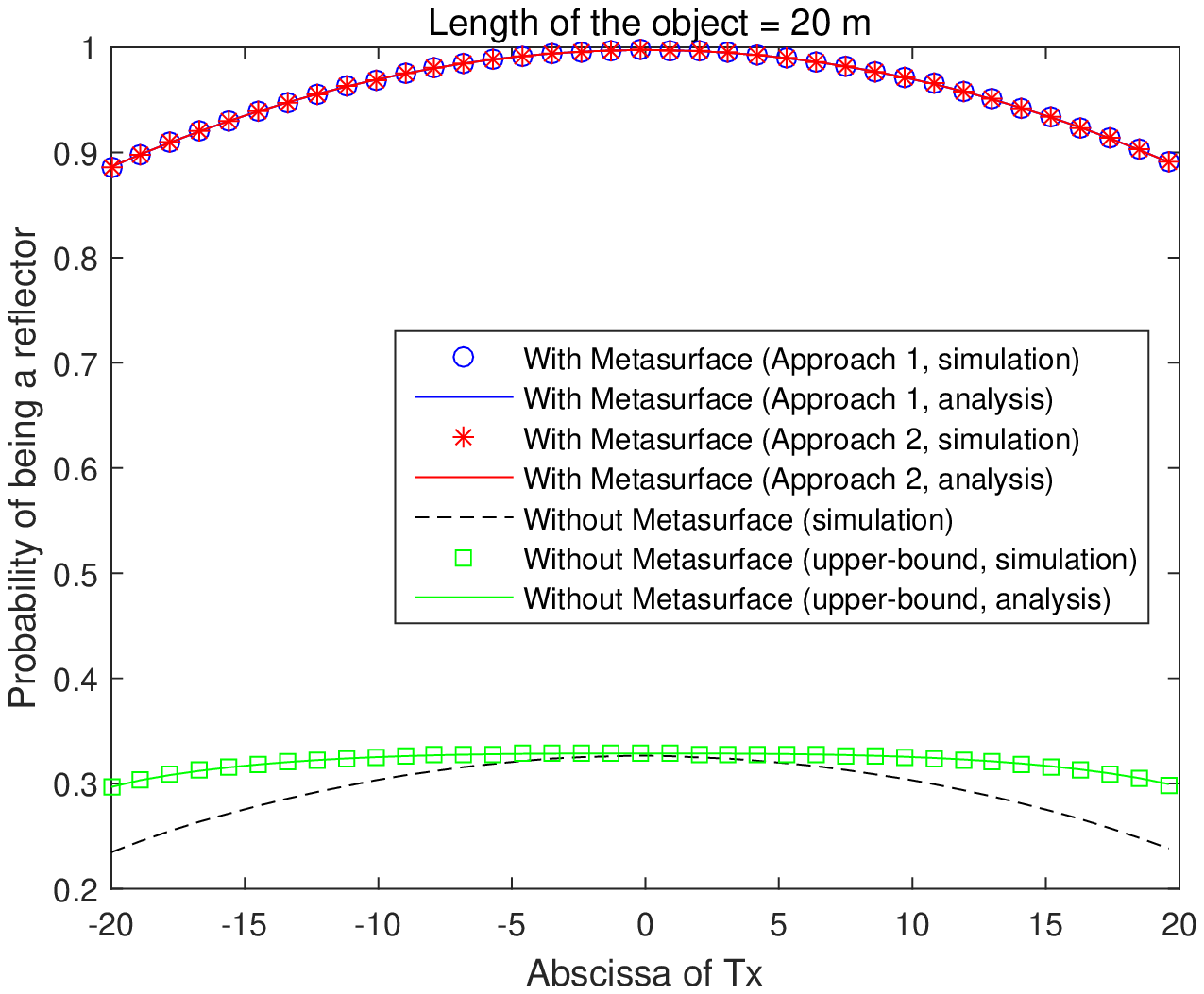}
	\caption{Probability of being a reflector versus the horizontal location, ${x_{{\rm{Tx}}}}$, of the transmitter. Setup: ${R_{{\rm{net}}}} = 30{\rm{ m}}$, vertical location of the transmitter ${y_{{\rm{Tx}}}} = 3$, location of the receiver $\left( {0,0} \right)$, length of the object $L = 20{\rm{ m}}$.} \label{Fig_10}
\end{figure}

The aim of this section is to validate the analytical frameworks developed in the previous sections against Monte Carlo simulations, and to study the potential of using reconfigurable metasurfaces in wireless networks. The results are illustrated either as a function of the length, $L$, of the typical object or as a function of the locations of the transmitter and receiver. The simulation setup is detailed in the caption of each figure. \\

In Fig. \ref{Fig_8}, we validate the proposed mathematical frameworks, against Monte Carlo simulations, as a function of the length, $L$, of the typical object. The results depicted in Fig. \ref{Fig_8} confirm the good accuracy of the proposed analytical approach. More importantly, we observe the large gain that the presence of metasurfaces bring about: Especially for objects of small length, the presence of reconfigurable metasurfaces increases the probability of the typical object to be a reflector significantly. This is expected to bring major gains in terms of signal strength of the received signal thanks to the reflection generated by the randomly distributed reflectors. The presence of multiple reflectors, however, may also increase the level of interference. Therefore, the optimization of wireless networks in the presence of reconfigurable metasurfaces is a challenging and open research issue. Figure \ref{Fig_8}, in addition, confirms the main finding in Remark \ref{remarkMDR}. \\

In Figs. \ref{Fig_9} and \ref{Fig_10}, we depict the probability that the typical object is as a reflector as a function of the location of the transmitter and for a fixed length of the typical object. Once again, the proposed analytical frameworks are accurate, and the upper-bound in Theorem \ref{theorem3} is sufficiently accurate for the considered setups. Especially for small-size objects, we observe the large gains that employing metasurfaces bring about. Even small-size objects can provide one with a relatively high probability of being a reflector, which is useful information in order to reduce the deployment cost of the metasurfaces over large-size environmental objects. Small-size metasurfaces, in fact, may be moved along large(size surfaces, and their location may be optimized in order to optimize the system performance. \\
\section{Conclusion and Discussion} \label{Conclusion} 
In this paper, we have proposed the first analytical approach that provides one with the probability that a random object coated with reconfigurable metasurfaces acts as a reflector, and have compared it against the conventional setup in which the object is not coated with reconfigurable metasurfaces. This result has been obtained by modeling the environmental objects with a modified random line process with fixed length, and random orientations and locations. Our proposed analytical approach allows us to prove that the probability that an object is a reflector does not depend on the length of the object if it is coated with metasurfaces, while it strongly depends on it if the Snell's law of reflection needs to be applied. The reason of this major difference in system performance lies in the fact that the angles of incidence and reflection need to be the same according to the Snell's law of reflection. \\

In spite of the novelty and contribution of the present paper, it constitutes only a first attempt to quantify the potential of reconfigurable metasurfaces in large-scale wireless networks, and to develop a general analytical approach for understanding the ultimate performance limits, and to identify design guidelines for system optimization. For example, the performance trends are based on the assumption that, for any angle of incidence, an arbitrary angle of reflection can be synthetized. Due to practical constraints on implementing metasurfaces, only a finite subset of angles may be allowed, which needs to account for the concept of field-of-view of the metasurfaces. Also, the analytical models and the simulation results have been obtained by using ray tracing assumptions, and ignore, e.g., the radiation pattern of the metasurfaces, and near field effects. A major step is needed to obtain tractable analytical expressions of relevant performance metrics that are suitable to unveil scaling laws, amenable for optimization, and account for different functions applied by the metasurfaces (not just reflections). \\

\appendix
\section{Proof of Lemma \ref{lemma3}}
Let us rewrite \eqref{eq6} and \eqref{eq9} in the following standard forms:
\begin{equation} \label{eq18}
\centering
\begin{gathered}
mx - y + z = 0 \hfill \\
{m_p} - y + {z_p} = 0 \hfill \\
\end{gathered}
\end{equation}
where their slopes $m$ and ${m_p}$ are assumed to be non-zero. \\

With this formulation, the directional vector of each line is as follows:
\begin{equation}
\centering
{\Delta} = \left[ {\begin{array}{*{20}{c}}
	m \\
	{ - 1}
	\end{array}} \right],{\rm{ }}{\Delta _p} = \left[ {\begin{array}{*{20}{c}}
	{{m_p}} \\
	{ - 1}
	\end{array}} \right]
\end{equation}

The two lines in \eqref{eq18} are perpendicular to each other if their directional vectors are orthogonal:
\begin{equation}
\centering
\Delta ^T{\Delta _p} = 0
\end{equation}
which implies that the following identity need to be fulfilled $m{m_p} =  - 1$. \\

The coordinates of the mid-point of the line segment between the transmitter and receiver is $\left( {\frac{{{x_{{\rm{Tx}}}} + {x_{{\rm{Rx}}}}}}{2},\frac{{{y_{{\rm{Tx}}}} + {y_{{\rm{Rx}}}}}}{2}} \right)$, which is located on the line $y = {m_p}x + {z_p}$. By substituting this latter point in \eqref{eq18}, ${z_p}$ turns out to be the following:
\begin{equation}
\centering
{z_p} = \frac{1}{{2m}}\left( {{x_{{\rm{Tx}}}} + {x_{{\rm{Rx}}}}} \right) + \frac{1}{2}\left( {{y_{{\rm{Tx}}}} + {y_{{\rm{Rx}}}}} \right)
\end{equation}
This concludes the proof. \\

\begin{figure}[!t]
	\setlength{\captionmargin}{10.0pt}
	\centering
	\includegraphics[width=0.50\columnwidth]{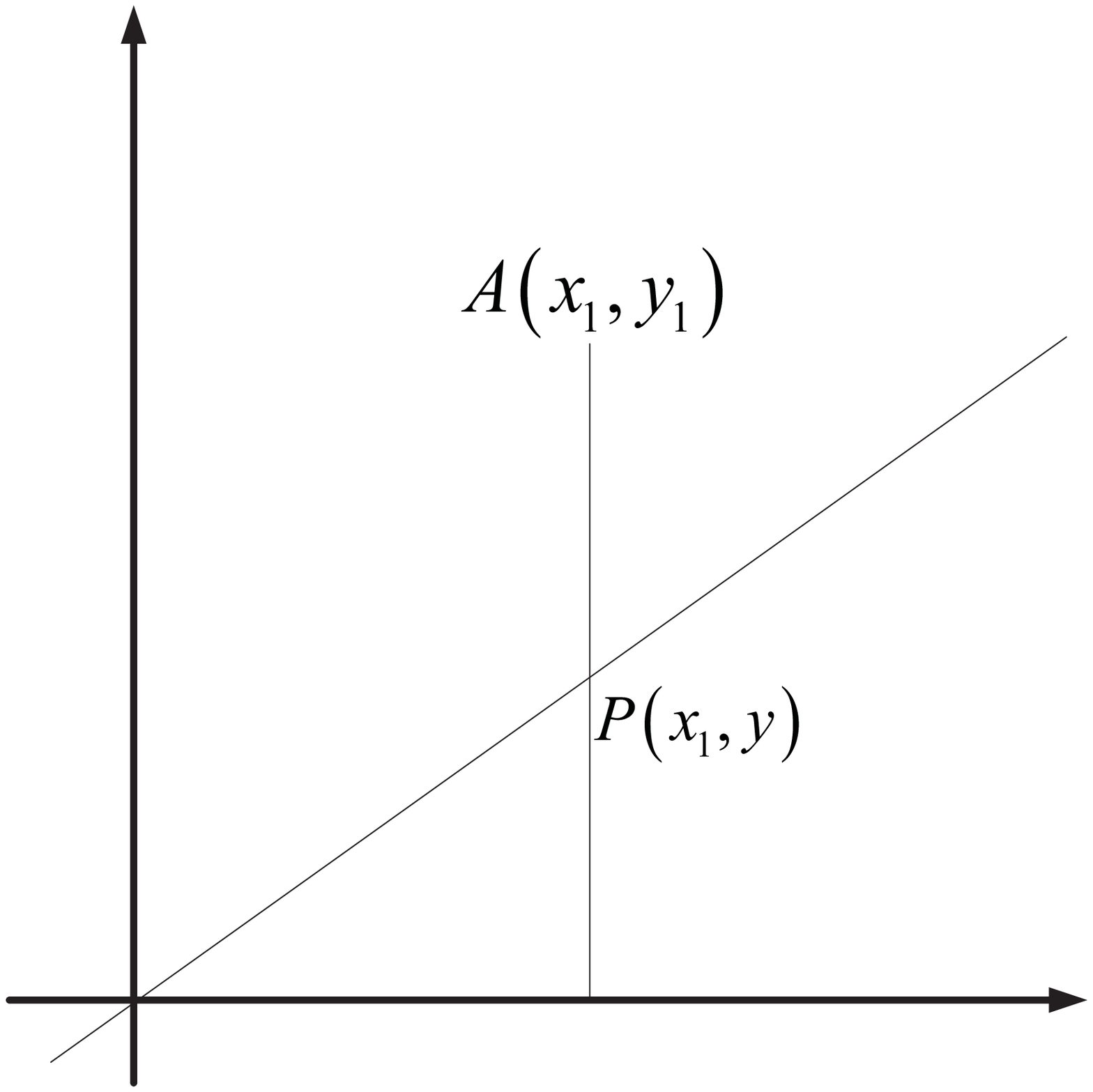}
	\caption{Position of a point with respect to a line.} \label{Fig_15}
\end{figure}
\section{Proof of Lemma \ref{lemma5}}
Let $A\left( {{x_1},{y_1}} \right)$ be a point that does not lie on a generic line $ax + by + c = 0$. Let us draw the perpendicular line from $A\left( {{x_1},{y_1}} \right)$ to the horizontal axis. This perpendicular line intersect the generic line $ax + by + c = 0$ in the point $P\left( {{x_1},{y}} \right)$, as shown in Fig. \ref{Fig_15}. By construction, ${{x_1}}$ is the abscissa of the point $P$. Let $y$ denote the ordinate of $P$. Since $P$ is on the line $ax + by + c = 0$, then the following equation needs to be satisfied:
\begin{equation}
\centering
a{x_1} + by + c = 0 \Rightarrow y =  - \frac{{a{x_1} + c}}{b}
\end{equation}

Let us compute the difference ${y_1} - y$, as follows:
\begin{equation} \label{eq22}
{y_1} - y = {y_1} - \left( { - \frac{{a{x_1} + c}}{b}} \right) \Rightarrow {y_1} - y = \frac{{a{x_1} + b{y_1} + c}}{b} \\
\end{equation}

Therefore, the following conclusions can be drawn: \\
\begin{itemize}
\item If the point $A$ is above the line, then it must be ${y_1} - y > 0$. From equation \eqref{eq22}, we evince that ${y_1} - y > 0$ if $\frac{{a{x_1} + b{y_1} + c}}{b} > 0$. This, in turn, corresponds to the following: i) either $a{x_1} + b{y_1} + c > 0$ and $b>0$ or ii) $a{x_1} + b{y_1} + c < 0$ and $b<0$.

\item If the point $A$ is below the line, then it must be ${y_1} - y < 0$. From equation \eqref{eq22}, we evince that ${y_1} - y < 0$ if $\frac{{a{x_1} + b{y_1} + c}}{b} < 0$. This, in turn, corresponds to the following: i) either $a{x_1} + b{y_1} + c < 0$ and $b>0$ or ii) $a{x_1} + b{y_1} + c > 0$ and $b<0$. \\
\end{itemize}

This concludes the proof. \\

\section{Proof of Theorem \ref{theorem1}}
From \eqref{eq10} in Lemma \ref{lemma4}, the probability of the complement of Event \ref{Event1} can be formulated as follows:
\begin{equation}
\centering
\begin{gathered}
\Pr \left\{ {\overline {{\rm{Event \; 1}}} } \right\} = \Pr \left\{ \begin{gathered}
\min \left( {{x_{{\rm{Tx}}}},{x_{{\rm{Rx}}}}} \right) \leqslant {x^*} \leqslant \max \left( {{x_{{\rm{Tx}}}},{x_{{\rm{Rx}}}}} \right){\rm{ }} \hfill \\
\cap {\rm{ }}\min \left( {{y_{{\rm{Tx}}}},{y_{{\rm{Rx}}}}} \right) \leqslant {y^*} \leqslant \max \left( {{y_{{\rm{Tx}}}},{y_{{\rm{Rx}}}}} \right) \hfill \\
\end{gathered}  \right\} \hfill \\
= \Pr \left\{ \begin{gathered}
\min \left( {{x_{{\rm{Tx}}}},{x_{{\rm{Rx}}}}} \right) \leqslant \frac{{p - z\sin \alpha }}{{m\sin \alpha  + \cos \alpha }} \leqslant \max \left( {{x_{{\rm{Tx}}}},{x_{{\rm{Rx}}}}} \right){\rm{ }} \hfill \\
\cap {\rm{ }}\min \left( {{y_{{\rm{Tx}}}},{y_{{\rm{Rx}}}}} \right) \leqslant m\frac{{p - z\sin \alpha }}{{m\sin \alpha  + \cos \alpha }} + z \leqslant \max \left( {{y_{{\rm{Tx}}}},{y_{{\rm{Rx}}}}} \right) \hfill \\
\end{gathered}  \right\} \hfill \\
\end{gathered}
\end{equation}

Based on the sign of ${m\sin \alpha  + \cos \alpha }$ and $m$, four cases can be identified.

\subsection{Case 1}
If $m\sin \alpha  + \cos \alpha  \geqslant 0$ and $m \geqslant 0$, we obtain the following:
\begin{equation}
\centering
\begin{gathered}
\Pr \left\{ {\overline {{\rm{Event \; 1}}} } \right\} = \Pr \left\{ \begin{gathered}
\min \left( {{x_{{\rm{Tx}}}},{x_{{\rm{Rx}}}}} \right) \leqslant \frac{{p - z\sin \alpha }}{{m\sin \alpha  + \cos \alpha }} \leqslant \max \left( {{x_{{\rm{Tx}}}},{x_{{\rm{Rx}}}}} \right){\rm{ }} \hfill \\
\cap {\rm{ }}\min \left( {{y_{{\rm{Tx}}}},{y_{{\rm{Rx}}}}} \right) \leqslant m\frac{{p - z\sin \alpha }}{{m\sin \alpha  + \cos \alpha }} + z \leqslant \max \left( {{y_{{\rm{Tx}}}},{y_{{\rm{Rx}}}}} \right) \hfill \\
\end{gathered}  \right\} \hfill \\
 \mathop  = \limits^{\left( a \right)} \Pr \left\{ \begin{gathered}
f\left( {\alpha ,\min \left( {{x_{{\rm{Tx}}}},{x_{{\rm{Rx}}}}} \right)} \right) \leqslant \upsilon  \leqslant f\left( {\alpha ,\max \left( {{x_{{\rm{Tx}}}},{x_{{\rm{Rx}}}}} \right)} \right){\rm{ }} \hfill \\
\cap {\rm{ }}g\left( {\alpha ,\min \left( {{y_{{\rm{Tx}}}},{y_{{\rm{Rx}}}}} \right)} \right) \leqslant \upsilon  \leqslant g\left( {\alpha ,\max \left( {{y_{{\rm{Tx}}}},{y_{{\rm{Rx}}}}} \right)} \right) \hfill \\
\end{gathered}  \right\} \hfill \\
= \frac{1}{{2\pi }}\int_0^{2\pi } {\int_{\max \left\{ {f\left( {\alpha ,\min \left( {{x_{{\rm{Tx}}}},{x_{{\rm{Rx}}}}} \right)} \right),g\left( {\alpha ,\min \left( {{y_{{\rm{Tx}}}},{y_{{\rm{Rx}}}}} \right)} \right),0} \right\}}^{\min \left\{ {f\left( {\alpha ,\max \left( {{x_{{\rm{Tx}}}},{x_{{\rm{Rx}}}}} \right)} \right),g\left( {\alpha ,\max \left( {{y_{{\rm{Tx}}}},{y_{{\rm{Rx}}}}} \right)} \right),1} \right\}} {{f_\upsilon }\left( \upsilon  \right)} }  \times H\left( {m\sin \alpha  + \cos \alpha } \right)H\left( m \right) \hfill \\
\times H\left( \begin{gathered}
\min \left\{ {f\left( {\alpha ,\max \left( {{x_{{\rm{Tx}}}},{x_{{\rm{Rx}}}}} \right)} \right),g\left( {\alpha ,\max \left( {{y_{{\rm{Tx}}}},{y_{{\rm{Rx}}}}} \right)} \right),1} \right\} \hfill \\
- \max \left\{ {f\left( {\alpha ,\min \left( {{x_{{\rm{Tx}}}},{x_{{\rm{Rx}}}}} \right)} \right),g\left( {\alpha ,\min \left( {{y_{{\rm{Tx}}}},{y_{{\rm{Rx}}}}} \right)} \right),0} \right\} \hfill \\
\end{gathered}  \right)d\upsilon d\alpha  \hfill \\
\mathop  = \limits^{\left( b \right)} \frac{1}{{2\pi }}\left\{ {\int_0^{{\delta _1}} {{\theta _1}\left( {\alpha ,{x_{{\rm{Tx}}}},{x_{{\rm{Rx}}}},{y_{{\rm{Tx}}}},{y_{{\rm{Rx}}}}} \right)} d\alpha  + \int_{{\delta _2}}^{2\pi } {{\theta _1}\left( {\alpha ,{x_{{\rm{Tx}}}},{x_{{\rm{Rx}}}},{y_{{\rm{Tx}}}},{y_{{\rm{Rx}}}}} \right)} d\alpha } \right\} \hfill \\
\end{gathered}
\end{equation}
where (a) follows from $p = {R_{{\rm{net}}}}\sqrt u  = {R_{{\rm{net}}}}\upsilon $, and (b) follows by computing the integral with respect to $\upsilon$, whose probability density function is ${f_\upsilon }\left( \upsilon  \right) = 2\upsilon $, since $u$ is uniformly distributed in $\left[ {0,1} \right]$. \\

The integration limits in (b) are determined from the conditions $m\sin \alpha  + \cos \alpha  \geqslant 0$, which implies $0 \leqslant \alpha  \leqslant {\delta _1}$ and ${\delta _2} \leqslant \alpha  \leqslant 2\pi $, where ${\delta _1} = 2{\tan ^{ - 1}}\left( {m + \sqrt {1 + {m^2}} } \right)$ and ${\delta _2} = 2\pi  + 2{\tan ^{ - 1}}\left( {m - \sqrt {1 + {m^2}} } \right)$. \\

The other three case studies can be obtained by using the same approach as for Case 1. Thus, the details are omitted and only the final result is reported. \\

\subsection{Case 2}
If $m\sin \alpha  + \cos \alpha  \geqslant 0$ and $m < 0$, we have the following:
\begin{equation}
\centering
\begin{gathered}
\Pr \left\{ {\overline {{\rm{Event \; 1}}} } \right\} = \Pr \left\{ \begin{gathered}
\min \left( {{x_{{\rm{Tx}}}},{x_{{\rm{Rx}}}}} \right) \leqslant \frac{{p - z\sin \alpha }}{{m\sin \alpha  + \cos \alpha }} \leqslant \max \left( {{x_{{\rm{Tx}}}},{x_{{\rm{Rx}}}}} \right){\rm{ }} \hfill \\
\cap {\rm{ }}\min \left( {{y_{{\rm{Tx}}}},{y_{{\rm{Rx}}}}} \right) \leqslant m\frac{{p - z\sin \alpha }}{{m\sin \alpha  + \cos \alpha }} + z \leqslant \max \left( {{y_{{\rm{Tx}}}},{y_{{\rm{Rx}}}}} \right) \hfill \\
\end{gathered}  \right\} \hfill \\
= \frac{1}{{2\pi }}\left\{ {\int_0^{{\delta _1}} {{\theta _2}\left( {\alpha ,{x_{{\rm{Tx}}}},{x_{{\rm{Rx}}}},{y_{{\rm{Tx}}}},{y_{{\rm{Rx}}}}} \right)} d\alpha  + \int_{{\delta _2}}^{2\pi } {{\theta _2}\left( {\alpha ,{x_{{\rm{Tx}}}},{x_{{\rm{Rx}}}},{y_{{\rm{Tx}}}},{y_{{\rm{Rx}}}}} \right)} d\alpha } \right\} \hfill \\
\end{gathered}
\end{equation}

\subsection{Case 3}
If $m\sin \alpha  + \cos \alpha  < 0$ and $m \geqslant 0$, we have the following:
\begin{equation}
\centering
\begin{gathered}
\Pr \left\{ {\overline {{\rm{Event \; 1}}} } \right\} = \Pr \left\{ \begin{gathered}
\min \left( {{x_{{\rm{Tx}}}},{x_{{\rm{Rx}}}}} \right) \leqslant \frac{{p - z\sin \alpha }}{{m\sin \alpha  + \cos \alpha }} \leqslant \max \left( {{x_{{\rm{Tx}}}},{x_{{\rm{Rx}}}}} \right){\rm{ }} \hfill \\
\cap {\rm{ }}\min \left( {{y_{{\rm{Tx}}}},{y_{{\rm{Rx}}}}} \right) \leqslant m\frac{{p - z\sin \alpha }}{{m\sin \alpha  + \cos \alpha }} + z \leqslant \max \left( {{y_{{\rm{Tx}}}},{y_{{\rm{Rx}}}}} \right) \hfill \\
\end{gathered}  \right\} \hfill \\
= \frac{1}{{2\pi }}\left\{ {\int_{{\delta _1}}^\pi  {{\theta _3}\left( {\alpha ,{x_{{\rm{Tx}}}},{x_{{\rm{Rx}}}},{y_{{\rm{Tx}}}},{y_{{\rm{Rx}}}}} \right)} d\alpha  + \int_\pi ^{{\delta _2}} {{\theta _3}\left( {\alpha ,{x_{{\rm{Tx}}}},{x_{{\rm{Rx}}}},{y_{{\rm{Tx}}}},{y_{{\rm{Rx}}}}} \right)} d\alpha } \right\} \hfill \\
= \frac{1}{{2\pi }}\int_{{\delta _1}}^{{\delta _2}} {{\theta _3}\left( {\alpha ,{x_{{\rm{Tx}}}},{x_{{\rm{Rx}}}},{y_{{\rm{Tx}}}},{y_{{\rm{Rx}}}}} \right)} d\alpha  \hfill \\
\end{gathered}
\end{equation}

\subsection{Case 4}
If $m\sin \alpha  + \cos \alpha  < 0$ and $m < 0$, we have the following:
\begin{equation}
\centering
\begin{gathered}
\Pr \left\{ {\overline {{\rm{Event \; 1}}} } \right\} = \Pr \left\{ \begin{gathered}
\min \left( {{x_{{\rm{Tx}}}},{x_{{\rm{Rx}}}}} \right) \leqslant \frac{{p - z\sin \alpha }}{{m\sin \alpha  + \cos \alpha }} \leqslant \max \left( {{x_{{\rm{Tx}}}},{x_{{\rm{Rx}}}}} \right){\rm{ }} \hfill \\
\cap {\rm{ }}\min \left( {{y_{{\rm{Tx}}}},{y_{{\rm{Rx}}}}} \right) \leqslant m\frac{{p - z\sin \alpha }}{{m\sin \alpha  + \cos \alpha }} + z \leqslant \max \left( {{y_{{\rm{Tx}}}},{y_{{\rm{Rx}}}}} \right) \hfill \\
\end{gathered}  \right\} \hfill \\
= \frac{1}{{2\pi }}\left\{ {\int_{{\delta _1}}^\pi  {{\theta _4}\left( {\alpha ,{x_{{\rm{Tx}}}},{x_{{\rm{Rx}}}},{y_{{\rm{Tx}}}},{y_{{\rm{Rx}}}}} \right)} d\alpha  + \int_\pi ^{{\delta _2}} {{\theta _4}\left( {\alpha ,{x_{{\rm{Tx}}}},{x_{{\rm{Rx}}}},{y_{{\rm{Tx}}}},{y_{{\rm{Rx}}}}} \right)} d\alpha } \right\} \hfill \\
= \frac{1}{{2\pi }}\int_{{\delta _1}}^{{\delta _2}} {{\theta _4}\left( {\alpha ,{x_{{\rm{Tx}}}},{x_{{\rm{Rx}}}},{y_{{\rm{Tx}}}},{y_{{\rm{Rx}}}}} \right)} d\alpha  \hfill \\
\end{gathered}
\end{equation}

This concludes the proof. \\

\section{Proof of Theorem \ref{theorem2}}
The proof is based on Lemma \ref{lemma5}. In particular, the following cases need to be examined. \\
\begin{itemize}
\item Case 1: The location of the transmitter $\left( {{x_{{\rm{Tx}}}},{y_{{\rm{Tx}}}}} \right)$ is above the line $x\cos \alpha  + y\sin \alpha  - p = 0$ given $\sin \alpha  > 0$.

\item Case 2: The location of the transmitter $\left( {{x_{{\rm{Tx}}}},{y_{{\rm{Tx}}}}} \right)$ is above the line $x\cos \alpha  + y\sin \alpha  - p = 0$ given $\sin \alpha  < 0$.

\item Case 3: The location of the receiver $\left( {{x_{{\rm{Rx}}}},{y_{{\rm{Rx}}}}} \right)$ is above the line $x\cos \alpha  + y\sin \alpha  - p = 0$ given $\sin \alpha  > 0$.

\item Case 4: The location of the receiver $\left( {{x_{{\rm{Rx}}}},{y_{{\rm{Rx}}}}} \right)$ is above the line $x\cos \alpha  + y\sin \alpha  - p = 0$ given $\sin \alpha  < 0$.

\item Case 5: The location of the transmitter $\left( {{x_{{\rm{Tx}}}},{y_{{\rm{Tx}}}}} \right)$ is below the line $x\cos \alpha  + y\sin \alpha  - p = 0$ given $\sin \alpha  > 0$.

\item Case 6: The location of the transmitter $\left( {{x_{{\rm{Tx}}}},{y_{{\rm{Tx}}}}} \right)$ is below the line $x\cos \alpha  + y\sin \alpha  - p = 0$ given $\sin \alpha  < 0$.

\item Case 7: The location of the receiver $\left( {{x_{{\rm{Rx}}}},{y_{{\rm{Rx}}}}} \right)$ is below the line $x\cos \alpha  + y\sin \alpha  - p = 0$ given $\sin \alpha  > 0$.

\item Case 8: The location of the receiver $\left( {{x_{{\rm{Rx}}}},{y_{{\rm{Rx}}}}} \right)$ is below the line $x\cos \alpha  + y\sin \alpha  - p = 0$ given $\sin \alpha  < 0$. \\
\end{itemize}

From \eqref{eq3}, the probability of Event \ref{Event1} can be formulated as follows:
\begin{equation}
\centering
\begin{gathered}
\Pr \left\{ {{\rm{Event \; 1}}} \right\} = \Pr \left\{ \begin{gathered}
\left\{ {\left[ {{\text{Tx is above the line}}} \right]{\rm{ }} \cap {\rm{ }}\left[ {{\text{Rx is above the line}}} \right]} \right\} \hfill \\
\cup \left\{ {\left[ {{\text{Tx is below the line}}} \right]{\rm{ }} \cap {\rm{ }}\left[ {{\text{Rx is below the line}}} \right]} \right\} \hfill \\
\end{gathered}  \right\} \hfill \\
= \Pr \left\{ \begin{gathered}
\left[ {{\text{Case 1 }} \cup {\text{ Case 2}}} \right]{\rm{ }} \cap {\rm{ }}\left[ {{\text{Case 3 }} \cup {\text{ Case 4}}} \right] \hfill \\
\cup {\rm{ }}\left[ {{\text{Case 5 }} \cup {\text{ Case 6}}} \right]{\rm{ }} \cap {\rm{ }}\left[ {{\text{Case 7 }} \cup {\text{ Case 8}}} \right] \hfill \\
\end{gathered}  \right\} \hfill \\
= \Pr \left\{ {\left[ {{\text{Case 1 }} \cup {\text{ Case 2}}} \right]{\rm{ }} \cap {\rm{ }}\left[ {{\text{Case 3 }} \cup {\text{ Case 4}}} \right]} \right\} \hfill \\
+ \Pr \left\{ {\left[ {{\text{Case 5 }} \cup {\text{ Case 6}}} \right]{\rm{ }} \cap {\rm{ }}\left[ {{\text{Case 7 }} \cup {\text{ Case 8}}} \right]} \right\} \hfill \\
= \Pr \left\{ {{\text{Case 1 }} \cap {\text{ Case 3}}} \right\} + \Pr \left\{ {{\text{Case 2 }} \cap {\text{ Case 4}}} \right\} \hfill \\
+ \Pr \left\{ {{\text{Case 5 }} \cap {\text{ Case 7}}} \right\} + \Pr \left\{ {{\text{Case 6 }} \cap {\text{ Case 8}}} \right\} \hfill \\
\end{gathered}
\end{equation}

Therefore, four probabilities need to be computed. Let us start with the first one:
\begin{equation}
\centering
\begin{gathered}
\Pr \left\{ {{\text{Case 1 }} \cap {\rm{ Case 3}}} \right\} \hfill \\
= \Pr \left\{ {\left( {{x_{{\rm{Tx}}}}\cos \alpha  + {y_{{\rm{Tx}}}}\sin \alpha  - p > 0{\rm{ }} \cap {\rm{ }}\sin \alpha  > 0} \right){\rm{ }} \cap {\rm{ }}\left( {{x_{{\rm{Rx}}}}\cos \alpha  + {y_{{\rm{Rx}}}}\sin \alpha  - p > 0{\rm{ }} \cap {\rm{ }}\sin \alpha  > 0} \right)} \right\} \hfill \\
= \Pr \left\{ {\left[ {\left( {{x_{{\rm{Tx}}}}\cos \alpha  + {y_{{\rm{Tx}}}}\sin \alpha  - p > 0} \right){\rm{ }} \cap {\rm{ }}\left( {{x_{{\rm{Rx}}}}\cos \alpha  + {y_{{\rm{Rx}}}}\sin \alpha  - p > 0} \right)} \right]{\rm{ }} \cap {\rm{ }}\sin \alpha  > 0} \right\} \hfill \\
\mathop  = \limits^{\left( a \right)} \Pr \left\{ {\left[ {\left( {\upsilon  < \frac{{{x_{{\rm{Tx}}}}\cos \alpha  + {y_{{\rm{Tx}}}}\sin \alpha }}{{{R_{{\rm{net}}}}}}} \right){\rm{ }} \cap {\rm{ }}\left( {\upsilon  < \frac{{{x_{{\rm{Rx}}}}\cos \alpha  + {y_{{\rm{Rx}}}}\sin \alpha }}{{{R_{{\rm{net}}}}}}} \right)} \right]{\rm{ }} \cap {\rm{ }}\sin \alpha  > 0} \right\} \hfill \\
= \frac{1}{{2\pi }}\int_0^\pi  {\int_0^{\min \left\{ {\frac{{{x_{{\rm{Tx}}}}\cos \alpha  + {y_{{\rm{Tx}}}}\sin \alpha }}{{{R_{{\rm{net}}}}}},\frac{{{x_{{\rm{Rx}}}}\cos \alpha  + {y_{{\rm{Rx}}}}\sin \alpha }}{{{R_{{\rm{net}}}}}},1} \right\}} {{f_\upsilon }\left( \upsilon  \right)} }  \hfill \\
\times H\left( {\min \left\{ {\frac{{{x_{{\rm{Tx}}}}\cos \alpha  + {y_{{\rm{Tx}}}}\sin \alpha }}{{{R_{{\rm{net}}}}}},\frac{{{x_{{\rm{Rx}}}}\cos \alpha  + {y_{{\rm{Rx}}}}\sin \alpha }}{{{R_{{\rm{net}}}}}},1} \right\}} \right)d\upsilon d\alpha  \hfill \\
\mathop  = \limits^{\left( b \right)} \frac{1}{{2\pi }}\int_0^\pi  {{\rho _1}\left( {\alpha ,{x_{{\rm{Tx}}}},{y_{{\rm{Tx}}}},{x_{{\rm{Rx}}}},{y_{{\rm{Rx}}}}} \right)} d\alpha  \hfill \\
\end{gathered}
\end{equation}
where (a) follows from $p = {R_{{\rm{net}}}}\sqrt u  = {R_{{\rm{net}}}}\upsilon $, and (b) follows by solving the integral with respect to $\upsilon$ whose probability density function is ${f_\upsilon }\left( \upsilon  \right) = 2\upsilon $, since $u$ is uniformly distributed in $\left[ {0,1} \right]$. \\

By using a similar approach, we we obtain the following results:
\begin{equation}
\centering
\begin{gathered}
\Pr \left\{ {{\text{Case 2 }} \cap {\text{ Case 4}}} \right\} = \frac{1}{{2\pi }}\int_\pi ^{2\pi } {{\rho _2}\left( {\alpha ,{x_{{\rm{Tx}}}},{y_{{\rm{Tx}}}},{x_{{\rm{Rx}}}},{y_{{\rm{Rx}}}}} \right)} d\alpha  \hfill \\
\Pr \left\{ {{\text{Case 5 }} \cap {\text{ Case 7}}} \right\} = \frac{1}{{2\pi }}\int_0^\pi  {{\rho _2}\left( {\alpha ,{x_{{\rm{Tx}}}},{y_{{\rm{Tx}}}},{x_{{\rm{Rx}}}},{y_{{\rm{Rx}}}}} \right)} d\alpha  \hfill \\
\Pr \left\{ {{\text{Case 6 }} \cap {\text{ Case 8}}} \right\} = \frac{1}{{2\pi }}\int_\pi ^{2\pi } {{\rho _1}\left( {\alpha ,{x_{{\rm{Tx}}}},{y_{{\rm{Tx}}}},{x_{{\rm{Rx}}}},{y_{{\rm{Rx}}}}} \right)} d\alpha  \hfill \\
\end{gathered}
\end{equation}

This concludes the proof.

\section{Proof of Proposition \ref{proposition1}}
From \eqref{eq8} in Lemma \ref{lemma2} and \eqref{eq11} in Lemma \ref{lemma4}, the probability of Event \ref{Event2} can be formulated as follows:
\begin{equation}
\centering
\begin{gathered}
\Pr \left\{ {{\text{Event 2}}} \right\} = \Pr \left\{ \begin{gathered}
\min \left( {{x_{{\rm{end1}}}},{x_{{\rm{end2}}}}} \right) \leqslant {x_*} \leqslant \max \left( {{x_{{\rm{end1}}}},{x_{{\rm{end2}}}}} \right){\rm{ }} \hfill \\
\cap {\rm{ }}\min \left( {{y_{{\rm{end1}}}},{y_{{\rm{end2}}}}} \right) \leqslant {y_*} \leqslant \max \left( {{y_{{\rm{end1}}}},{y_{{\rm{end2}}}}} \right) \hfill \\
\end{gathered}  \right\} \hfill \\
= \Pr \left\{ \begin{gathered}
\min \left( {{x_{{\rm{end1}}}},{x_{{\rm{end2}}}}} \right) \leqslant \frac{{p - {z_p}\sin \alpha }}{{{m_p}\sin \alpha  + \cos \alpha }} \leqslant \max \left( {{x_{{\rm{end1}}}},{x_{{\rm{end2}}}}} \right){\rm{ }} \hfill \\
\cap {\rm{ }}\min \left( {{y_{{\rm{end1}}}},{y_{{\rm{end2}}}}} \right) \leqslant {m_p}\frac{{p - {z_p}\sin \alpha }}{{{m_p}\sin \alpha  + \cos \alpha }} + {z_p} \leqslant \max \left( {{y_{{\rm{end1}}}},{y_{{\rm{end2}}}}} \right) \hfill \\
\end{gathered}  \right\} \hfill \\
\end{gathered}
\end{equation}

In order to compute this probability, we need to examine four cases depending on the relationship between ${{x_{{\rm{end1}}}}}$ and ${{x_{{\rm{end2}}}}}$, as well as ${{y_{{\rm{end1}}}}}$ and ${{y_{{\rm{end2}}}}}$. \\

\subsection{Case 1}
If ${x_{{\rm{end1}}}} > {x_{{\rm{end2}}}}$ and ${y_{{\rm{end1}}}} > {y_{{\rm{end2}}}}$, which implies $\sin \alpha  < 0$ and $\cos \alpha  > 0$, we obtain the following:
\begin{equation}
\centering
\small
\begin{gathered}
\Pr \left\{ {{\text{Event 2}}} \right\} = \Pr \left\{ {{x_{{\rm{end2}}}} \leqslant \frac{{p - {z_p}\sin \alpha }}{{{m_p}\sin \alpha  + \cos \alpha }} \leqslant {x_{{\rm{end1}}}}{\rm{ }} \cap {\rm{ }}{y_{{\rm{end2}}}} \leqslant {m_p}\frac{{p - {z_p}\sin \alpha }}{{{m_p}\sin \alpha  + \cos \alpha }} + {z_p} \leqslant {y_{{\rm{end1}}}}} \right\} \hfill \\
= \Pr \left\{ \begin{gathered}
\frac{L}{2}\sin \alpha  + \frac{{{z_p}\sin \alpha }}{{{m_p}\sin \alpha  + \cos \alpha }} \leqslant \left( {\frac{1}{{{m_p}\sin \alpha  + \cos \alpha }} - \cos \alpha } \right)p \leqslant  - \frac{L}{2}\sin \alpha  + \frac{{{z_p}\sin \alpha }}{{{m_p}\sin \alpha  + \cos \alpha }}{\rm{ }} \cap {\rm{ }} \hfill \\
- \frac{L}{2}\cos \alpha  + \frac{{{m_p}{z_p}\sin \alpha }}{{{m_p}\sin \alpha  + \cos \alpha }} - {z_p} \leqslant \left( {\frac{{{m_p}}}{{{m_p}\sin \alpha  + \cos \alpha }} - \sin \alpha } \right)p \leqslant \frac{L}{2}\cos \alpha  + \frac{{{m_p}{z_p}\sin \alpha }}{{{m_p}\sin \alpha  + \cos \alpha }} - {z_p} \hfill \\
\end{gathered}  \right\} \hfill \\
\end{gathered}
\end{equation}

Depending on the sign of $\left( {\frac{1}{{{m_p}\sin \alpha  + \cos \alpha }} - \cos \alpha } \right)$ and $\left( {\frac{{{m_p}}}{{{m_p}\sin \alpha  + \cos \alpha }} - \sin \alpha } \right)$, four sub-cases need to be studied.

\subsubsection{Case 1-a}
If $\left( {\frac{1}{{{m_p}\sin \alpha  + \cos \alpha }} - \cos \alpha } \right) > 0$ and $\left( {\frac{{{m_p}}}{{{m_p}\sin \alpha  + \cos \alpha }} - \sin \alpha } \right) > 0$, we have the following:
\begin{equation}
\small
\begin{gathered}
\Pr \left\{ {{\text{Event 2}}} \right\} = \Pr \left\{ {{x_{{\rm{end2}}}} \leqslant \frac{{p - {z_p}\sin \alpha }}{{{m_p}\sin \alpha  + \cos \alpha }} \leqslant {x_{{\rm{end1}}}}{\rm{ }} \cap {\rm{ }}{y_{{\rm{end2}}}} \leqslant {m_p}\frac{{p - {z_p}\sin \alpha }}{{{m_p}\sin \alpha  + \cos \alpha }} + {z_p} \leqslant {y_{{\rm{end1}}}}} \right\} \hfill \\
\mathop  = \limits^{\left( a \right)} \Pr \left\{ {F\left( {\alpha ,\frac{L}{2}\sin \alpha } \right) \leqslant \upsilon  \leqslant F\left( {\alpha , - \frac{L}{2}\sin \alpha } \right){\rm{ }} \cap {\rm{ }}G\left( {\alpha , - \frac{L}{2}\cos \alpha } \right) \leqslant \upsilon  \leqslant G\left( {\alpha ,\frac{L}{2}\cos \alpha } \right)} \right\} \hfill \\
= \frac{1}{{2\pi }}\int_0^{2\pi } {\int_{\max \left\{ {F\left( {\alpha ,\frac{L}{2}\sin \alpha } \right),G\left( {\alpha , - \frac{L}{2}\cos \alpha } \right),0} \right\}}^{\min \left\{ {F\left( {\alpha , - \frac{L}{2}\sin \alpha } \right),F\left( {\alpha , - \frac{L}{2}\sin \alpha } \right),1} \right\}} {{f_\upsilon }\left( \upsilon  \right)} } H\left( {\frac{1}{{{m_p}\sin \alpha  + \cos \alpha }} - \cos \alpha } \right)H\left( {\frac{{{m_p}}}{{{m_p}\sin \alpha  + \cos \alpha }} - \sin \alpha } \right) \hfill \\
\times H\left( \begin{gathered}
\min \left\{ {F\left( {\alpha , - \frac{L}{2}\sin \alpha } \right),F\left( {\alpha , - \frac{L}{2}\sin \alpha } \right),1} \right\} \hfill \\
- \max \left\{ {F\left( {\alpha ,\frac{L}{2}\sin \alpha } \right),G\left( {\alpha , - \frac{L}{2}\cos \alpha } \right),0} \right\} \hfill \\
\end{gathered}  \right)\bar H\left( {\sin \alpha } \right)H\left( {\cos \alpha } \right)d\upsilon d\alpha  \hfill \\
\mathop  = \limits^{\left( b \right)} \frac{1}{{2\pi }}\int_0^{2\pi } {\Gamma _1^a\left( \alpha  \right)} \bar H\left( {\sin \alpha } \right)H\left( {\cos \alpha } \right)d\alpha  = \frac{1}{{2\pi }}\int_{\frac{{3\pi }}{2}}^{2\pi } {\Gamma _1^a\left( \alpha  \right)} d\alpha  \hfill \\
\end{gathered}
\end{equation}
where (a) follows from $p = {R_{{\rm{net}}}}\sqrt u  = {R_{{\rm{net}}}}\upsilon $, and (b) follows by solving the integral with respect to $\upsilon$ whose probability density function is ${f_\upsilon }\left( \upsilon  \right) = 2\upsilon $, since $u$ is uniformly distributed in $\left[ {0,1} \right]$. \\

By using a similar approach, we can study the remaining three sub-cases. \\

\subsubsection{Case 1-b}
If $\left( {\frac{1}{{{m_p}\sin \alpha  + \cos \alpha }} - \cos \alpha } \right) > 0$ and $\left( {\frac{{{m_p}}}{{{m_p}\sin \alpha  + \cos \alpha }} - \sin \alpha } \right) < 0$, we have the following:
\begin{equation}
\centering
\Pr \left\{ {{\text{Event 2}}} \right\} = \frac{1}{{2\pi }}\int_0^{2\pi } {\Gamma _1^b\left( \alpha  \right)} \bar H\left( {\sin \alpha } \right)H\left( {\cos \alpha } \right)d\alpha  = \frac{1}{{2\pi }}\int_{\frac{{3\pi }}{2}}^{2\pi } {\Gamma _1^b\left( \alpha  \right)} d\alpha
\end{equation}

\subsubsection{Case 1-c}
If $\left( {\frac{1}{{{m_p}\sin \alpha  + \cos \alpha }} - \cos \alpha } \right) < 0$ and $\left( {\frac{{{m_p}}}{{{m_p}\sin \alpha  + \cos \alpha }} - \sin \alpha } \right) > 0$, we have the following:
\begin{equation}
\centering
\Pr \left\{ {{\rm{Event 2}}} \right\} = \frac{1}{{2\pi }}\int_0^{2\pi } {\Gamma _1^c\left( \alpha  \right)} \bar H\left( {\sin \alpha } \right)H\left( {\cos \alpha } \right)d\alpha  = \frac{1}{{2\pi }}\int_{\frac{{3\pi }}{2}}^{2\pi } {\Gamma _1^c\left( \alpha  \right)} d\alpha
\end{equation}

\subsubsection{Case 1-d}
If $\left( {\frac{1}{{{m_p}\sin \alpha  + \cos \alpha }} - \cos \alpha } \right) < 0$ and $\left( {\frac{{{m_p}}}{{{m_p}\sin \alpha  + \cos \alpha }} - \sin \alpha } \right) < 0$, we have the following:
\begin{equation}
\centering
\Pr \left\{ {{\text{Event 2}}} \right\} = \frac{1}{{2\pi }}\int_0^{2\pi } {\Gamma _1^d\left( \alpha  \right)} \bar H\left( {\sin \alpha } \right)H\left( {\cos \alpha } \right)d\alpha  = \frac{1}{{2\pi }}\int_{\frac{{3\pi }}{2}}^{2\pi } {\Gamma _1^d\left( \alpha  \right)} d\alpha
\end{equation}

Therefore, eventually, $\Pr \left\{ {{\rm{Event 2}}} \right\}$ can be formulated as follows:
\begin{equation}
\centering
\Pr \left\{ {{\text{Event 2}}} \right\} = \frac{1}{{2\pi }}\int_{\frac{{3\pi }}{2}}^{2\pi } {{\Gamma _1}\left( \alpha  \right)} d\alpha  = \frac{1}{{2\pi }}\int_{\frac{{3\pi }}{2}}^{2\pi } {\left[ {\Gamma _1^a\left( \alpha  \right) + \Gamma _1^b\left( \alpha  \right) + \Gamma _1^c\left( \alpha  \right) + \Gamma _1^d\left( \alpha  \right)} \right]} d\alpha
\end{equation}

By using a similar line of thought, the remaining three cases can be studied. The final result is reported in the following sections. \\

\subsection{Case 2}
If ${x_{{\rm{end1}}}} > {x_{{\rm{end2}}}}$, and ${y_{{\rm{end1}}}} < {y_{{\rm{end2}}}}$, which implies $\sin \alpha  < 0$ and $\cos \alpha  < 0$, we obtain the following:
\begin{equation}
\centering
\Pr \left\{ {{\text{Event 2}}} \right\} = \frac{1}{{2\pi }}\int_\pi ^{\frac{{3\pi }}{2}} {{\Gamma _2}\left( \alpha  \right)} d\alpha  = \frac{1}{{2\pi }}\int_\pi ^{\frac{{3\pi }}{2}} {\left[ {\Gamma _2^a\left( \alpha  \right) + \Gamma _2^b\left( \alpha  \right) + \Gamma _2^c\left( \alpha  \right) + \Gamma _2^d\left( \alpha  \right)} \right]} d\alpha
\end{equation}

\subsection{Case 3}
If ${x_{{\rm{end1}}}} < {x_{{\rm{end2}}}}$, and ${y_{{\rm{end1}}}} > {y_{{\rm{end2}}}}$, which implies $\sin \alpha  > 0$ and $\cos \alpha  > 0$, we have the following:
\begin{equation}
\centering
\Pr \left\{ {{\text{Event 2}}} \right\} = \frac{1}{{2\pi }}\int_0^{\frac{\pi }{2}} {{\Gamma _3}\left( \alpha  \right)} d\alpha  = \frac{1}{{2\pi }}\int_0^{\frac{\pi }{2}} {\left[ {\Gamma _3^a\left( \alpha  \right) + \Gamma _3^b\left( \alpha  \right) + \Gamma _3^c\left( \alpha  \right) + \Gamma _3^d\left( \alpha  \right)} \right]} d\alpha
\end{equation}

\subsection{Case 4}
If ${x_{{\rm{end1}}}} < {x_{{\rm{end2}}}}$, and ${y_{{\rm{end1}}}} < {y_{{\rm{end2}}}}$, which implies $\sin \alpha  > 0$ and $\cos \alpha  < 0$, we have the following:
\begin{equation}
\centering
\Pr \left\{ {{\text{Event 2}}} \right\} = \frac{1}{{2\pi }}\int_0^{\frac{\pi }{2}} {{\Gamma _4}\left( \alpha  \right)} d\alpha  = \frac{1}{{2\pi }}\int_{\frac{\pi }{2}}^\pi  {\left[ {\Gamma _4^a\left( \alpha  \right) + \Gamma _4^b\left( \alpha  \right) + \Gamma _4^c\left( \alpha  \right) + \Gamma _4^d\left( \alpha  \right)} \right]} d\alpha
\end{equation}

This concludes the proof.


\begin{backmatter}

%
%
%
%

	
	\bibliographystyle{bmc-mathphys} 


\begin{thebibliography}{13}
\ifx \bisbn   \undefined \def \bisbn  #1{ISBN #1}\fi
\ifx \binits  \undefined \def \binits#1{#1}\fi
\ifx \bauthor  \undefined \def \bauthor#1{#1}\fi
\ifx \batitle  \undefined \def \batitle#1{#1}\fi
\ifx \bjtitle  \undefined \def \bjtitle#1{#1}\fi
\ifx \bvolume  \undefined \def \bvolume#1{\textbf{#1}}\fi
\ifx \byear  \undefined \def \byear#1{#1}\fi
\ifx \bissue  \undefined \def \bissue#1{#1}\fi
\ifx \bfpage  \undefined \def \bfpage#1{#1}\fi
\ifx \blpage  \undefined \def \blpage #1{#1}\fi
\ifx \burl  \undefined \def \burl#1{\textsf{#1}}\fi
\ifx \doiurl  \undefined \def \doiurl#1{\textsf{#1}}\fi
\ifx \betal  \undefined \def \betal{\textit{et al.}}\fi
\ifx \binstitute  \undefined \def \binstitute#1{#1}\fi
\ifx \binstitutionaled  \undefined \def \binstitutionaled#1{#1}\fi
\ifx \bctitle  \undefined \def \bctitle#1{#1}\fi
\ifx \beditor  \undefined \def \beditor#1{#1}\fi
\ifx \bpublisher  \undefined \def \bpublisher#1{#1}\fi
\ifx \bbtitle  \undefined \def \bbtitle#1{#1}\fi
\ifx \bedition  \undefined \def \bedition#1{#1}\fi
\ifx \bseriesno  \undefined \def \bseriesno#1{#1}\fi
\ifx \blocation  \undefined \def \blocation#1{#1}\fi
\ifx \bsertitle  \undefined \def \bsertitle#1{#1}\fi
\ifx \bsnm \undefined \def \bsnm#1{#1}\fi
\ifx \bsuffix \undefined \def \bsuffix#1{#1}\fi
\ifx \bparticle \undefined \def \bparticle#1{#1}\fi
\ifx \barticle \undefined \def \barticle#1{#1}\fi
\ifx \bconfdate \undefined \def \bconfdate #1{#1}\fi
\ifx \botherref \undefined \def \botherref #1{#1}\fi
\ifx \url \undefined \def \url#1{\textsf{#1}}\fi
\ifx \bchapter \undefined \def \bchapter#1{#1}\fi
\ifx \bbook \undefined \def \bbook#1{#1}\fi
\ifx \bcomment \undefined \def \bcomment#1{#1}\fi
\ifx \oauthor \undefined \def \oauthor#1{#1}\fi
\ifx \citeauthoryear \undefined \def \citeauthoryear#1{#1}\fi
\ifx \endbibitem  \undefined \def \endbibitem {}\fi
\ifx \bconflocation  \undefined \def \bconflocation#1{#1}\fi
\ifx \arxivurl  \undefined \def \arxivurl#1{\textsf{#1}}\fi
\csname PreBibitemsHook\endcsname

\bibitem{koon}
\begin{barticle}
\bauthor{\bsnm{Koonin}, \binits{E.V.}},
\bauthor{\bsnm{Altschul}, \binits{S.F.}},
\bauthor{\bsnm{Bork}, \binits{P.}}:
\batitle{Brca1 protein products: functional motifs}.
\bjtitle{Nat Genet}
\bvolume{13},
\bfpage{266}--\blpage{267}
(\byear{1996})
\end{barticle}
\endbibitem

\bibitem{khar}
\begin{botherref}
\oauthor{\bsnm{Kharitonov}, \binits{S.A.}},
\oauthor{\bsnm{Barnes}, \binits{P.J.}}:
Clinical Aspects of Exhaled Nitric Oxide.
in press
\end{botherref}
\endbibitem

\bibitem{zvai}
\begin{barticle}
\bauthor{\bsnm{Zvaifler}, \binits{N.J.}},
\bauthor{\bsnm{Burger}, \binits{J.A.}},
\bauthor{\bsnm{Marinova-Mutafchieva}, \binits{L.}},
\bauthor{\bsnm{Taylor}, \binits{P.}},
\bauthor{\bsnm{Maini}, \binits{R.N.}}:
\batitle{Mesenchymal cells, stromal derived factor-1 and rheumatoid arthritis
  [abstract]}.
\bjtitle{Arthritis Rheum}
\bvolume{42},
\bfpage{250}
(\byear{1999})
\end{barticle}
\endbibitem

\bibitem{xjon}
\begin{bchapter}
\bauthor{\bsnm{Jones}, \binits{X.}}:
\bctitle{Zeolites and synthetic mechanisms}.
In: \beditor{\bsnm{Smith}, \binits{Y.}} (ed.)
\bbtitle{Proceedings of the First National Conference on Porous Sieves: 27-30
  June 1996; Baltimore},
pp. \bfpage{16}--\blpage{27}
(\byear{1996}).
\bcomment{Stoneham: Butterworth-Heinemann}
\end{bchapter}
\endbibitem

\bibitem{marg}
\begin{bbook}
\bauthor{\bsnm{Margulis}, \binits{L.}}:
\bbtitle{Origin of Eukaryotic Cells}.
\bpublisher{Yale University Press},
\blocation{New Haven}
(\byear{1970})
\end{bbook}
\endbibitem

\bibitem{oreg}
\begin{barticle}
\bauthor{\bsnm{Orengo}, \binits{C.A.}},
\bauthor{\bsnm{Bray}, \binits{J.E.}},
\bauthor{\bsnm{Hubbard}, \binits{T.}},
\bauthor{\bsnm{LoConte}, \binits{L.}},
\bauthor{\bsnm{Sillitoe}, \binits{I.}}:
\batitle{Analysis and assessment of ab initio three-dimensional prediction,
  secondary structure, and contacts prediction}.
\bjtitle{Proteins}
\bvolume{Suppl 3},
\bfpage{149}--\blpage{170}
(\byear{1999})
\end{barticle}
\endbibitem

\bibitem{schn}
\begin{bchapter}
\bauthor{\bsnm{Schnepf}, \binits{E.}}:
\bctitle{From prey via endosymbiont to plastids: comparative studies in
  dinoflagellates}.
In: \beditor{\bsnm{Lewin}, \binits{R.A.}} (ed.)
\bbtitle{Origins of Plastids}
vol. \bseriesno{2},
\bedition{2nd} edn.,
pp. \bfpage{53}--\blpage{76}.
\bpublisher{Chapman and Hall},
\blocation{New York}
(\byear{1993})
\end{bchapter}
\endbibitem

\bibitem{pond}
\begin{botherref}
Innovative Oncology
\end{botherref}
\endbibitem

\bibitem{smith}
\begin{bbook}
\beditor{\bsnm{Smith}, \binits{Y.}} (ed.):
\bbtitle{Proceedings of the First National Conference on Porous Sieves: 27-30
  June 1996; Baltimore}.
\bpublisher{Butterworth-Heinemann},
\blocation{Stoneham}
(\byear{1996})
\end{bbook}
\endbibitem

\bibitem{hunn}
\begin{bchapter}
\bauthor{\bsnm{Hunninghake}, \binits{G.W.}},
\bauthor{\bsnm{Gadek}, \binits{J.E.}}:
\bctitle{The alveloar macrophage}.
In: \beditor{\bsnm{Harris}, \binits{T.J.R.}} (ed.)
\bbtitle{Cultured Human Cells and Tissues},
pp. \bfpage{54}--\blpage{56}.
\bpublisher{Academic Press},
\blocation{New York}
(\byear{1995}).
\bcomment{Stoner G (Series Editor): Methods and Perspectives in Cell Biology,
  vol 1}
\end{bchapter}
\endbibitem

\bibitem{advi}
\begin{botherref}
Advisory Committee on Genetic Modification:
Annual Report.
London
(1999).
Advisory Committee on Genetic Modification
\end{botherref}
\endbibitem

\bibitem{koha}
\begin{botherref}
\oauthor{\bsnm{Kohavi}, \binits{R.}}:
Wrappers for performance enhancement and obvious decision graphs.
PhD thesis,
Stanford University, Computer Science Department
(1995)
\end{botherref}
\endbibitem

\bibitem{mouse}
\begin{botherref}
The Mouse Tumor Biology Database.
\url{http://tumor.informatics.jax.org/cancer\_links.html}
\end{botherref}
\endbibitem

\end{thebibliography}

\newcommand{\BMCxmlcomment}[1]{}

\BMCxmlcomment{

<refgrp>

<bibl id="B1">
  <title><p>BRCA1 protein products: functional motifs</p></title>
  <aug>
    <au><snm>Koonin</snm><fnm>E V</fnm></au>
    <au><snm>Altschul</snm><fnm>S F</fnm></au>
    <au><snm>Bork</snm><fnm>P</fnm></au>
  </aug>
  <source>Nat Genet</source>
  <pubdate>1996</pubdate>
  <volume>13</volume>
  <fpage>266</fpage>
  <lpage>267</lpage>
</bibl>

<bibl id="B2">
  <title><p>Clinical aspects of exhaled nitric oxide</p></title>
  <aug>
    <au><snm>Kharitonov</snm><fnm>S A</fnm></au>
    <au><snm>Barnes</snm><fnm>P J</fnm></au>
  </aug>
  <source>Eur Respir J</source>
  <inpress />
</bibl>

<bibl id="B3">
  <title><p>Mesenchymal cells, stromal derived factor-1 and rheumatoid
  arthritis [abstract]</p></title>
  <aug>
    <au><snm>Zvaifler</snm><fnm>N J</fnm></au>
    <au><snm>Burger</snm><fnm>J A</fnm></au>
    <au><snm>Marinova Mutafchieva</snm><fnm>L</fnm></au>
    <au><snm>Taylor</snm><fnm>P</fnm></au>
    <au><snm>Maini</snm><fnm>R N</fnm></au>
  </aug>
  <source>Arthritis Rheum</source>
  <pubdate>1999</pubdate>
  <volume>42</volume>
  <fpage>s250</fpage>
</bibl>

<bibl id="B4">
  <title><p>Zeolites and synthetic mechanisms</p></title>
  <aug>
    <au><snm>Jones</snm><fnm>X</fnm></au>
  </aug>
  <source>Proceedings of the First National Conference on Porous Sieves: 27-30
  June 1996; Baltimore</source>
  <editor>Y Smith</editor>
  <pubdate>1996</pubdate>
  <fpage>16</fpage>
  <lpage>27</lpage>
</bibl>

<bibl id="B5">
  <title><p>Origin of Eukaryotic Cells</p></title>
  <aug>
    <au><snm>Margulis</snm><fnm>L</fnm></au>
  </aug>
  <publisher>New Haven: Yale University Press</publisher>
  <pubdate>1970</pubdate>
</bibl>

<bibl id="B6">
  <title><p>Analysis and assessment of ab initio three-dimensional prediction,
  secondary structure, and contacts prediction</p></title>
  <aug>
    <au><snm>Orengo</snm><fnm>C A</fnm></au>
    <au><snm>Bray</snm><fnm>J E</fnm></au>
    <au><snm>Hubbard</snm><fnm>T</fnm></au>
    <au><snm>LoConte</snm><fnm>L</fnm></au>
    <au><snm>Sillitoe</snm><fnm>I</fnm></au>
  </aug>
  <source>Proteins</source>
  <pubdate>1999</pubdate>
  <volume>Suppl 3</volume>
  <fpage>149</fpage>
  <lpage>170</lpage>
</bibl>

<bibl id="B7">
  <title><p>From prey via endosymbiont to plastids: comparative studies in
  dinoflagellates</p></title>
  <aug>
    <au><snm>Schnepf</snm><fnm>E</fnm></au>
  </aug>
  <source>Origins of Plastids</source>
  <publisher>New York: Chapman and Hall</publisher>
  <editor>R A Lewin</editor>
  <edition>2</edition>
  <pubdate>1993</pubdate>
  <volume>2</volume>
  <fpage>53</fpage>
  <lpage>76</lpage>
</bibl>

<bibl id="B8">
  <title><p>Innovative oncology</p></title>
  <source>Breast Cancer Res</source>
  <editor>B Ponder and S Johnston and L Chodosh</editor>
  <pubdate>1998</pubdate>
  <volume>10</volume>
  <fpage>1</fpage>
  <lpage>72</lpage>
</bibl>

<bibl id="B9">
  <title><p>Proceedings of the First National Conference on Porous Sieves:
  27-30 June 1996; Baltimore</p></title>
  <publisher>Stoneham: Butterworth-Heinemann</publisher>
  <editor>Y Smith</editor>
  <pubdate>1996</pubdate>
</bibl>

<bibl id="B10">
  <title><p>The alveloar macrophage</p></title>
  <aug>
    <au><snm>Hunninghake</snm><fnm>G W</fnm></au>
    <au><snm>Gadek</snm><fnm>J E</fnm></au>
  </aug>
  <source>Cultured Human Cells and Tissues</source>
  <publisher>New York: Academic Press</publisher>
  <editor>T J R Harris</editor>
  <pubdate>1995</pubdate>
  <fpage>54</fpage>
  <lpage>56</lpage>
  <note>Stoner G (Series Editor): Methods and Perspectives in Cell Biology, vol
  1</note>
</bibl>

<bibl id="B11">
  <title><p>Annual Report</p></title>
  <aug><au><cnm>Advisory Committee on Genetic Modification</cnm></au></aug>
  <publisher>London</publisher>
  <pubdate>1999</pubdate>
</bibl>

<bibl id="B12">
  <title><p>Wrappers for performance enhancement and obvious decision
  graphs</p></title>
  <aug>
    <au><snm>Kohavi</snm><fnm>R</fnm></au>
  </aug>
  <source>PhD thesis</source>
  <publisher>Stanford University, Computer Science Department</publisher>
  <pubdate>1995</pubdate>
</bibl>

<bibl id="B13">
  <title><p>The Mouse Tumor Biology Database</p></title>
  <url>http://tumor.informatics.jax.org/cancer\_links.html</url>
</bibl>

</refgrp>
} 


\begin{thebibliography}{99}

\bibitem{1} https://www.comsoc.org/ctn/what-will-6g-be.

\bibitem{2} P. Hu, P. Zhang, M. Rostami, and D. Ganesan, “Braidio: An Integrated Active-Passive Radio for Mobile Devices with Asymmetric Energy Budgets”, ACM SIGCOMM, Florianopolis, Brazil, Aug. 2016.

\bibitem{3} C. Liaskos, S. Nie, A. Tsioliaridou, A. Pitsillides, S. Ioannidis, and I. F. Akyildiz, “Realizing Wireless Communication Through Software-Defined HyperSurface Environments”, IEEE International Symposium on a World of Wireless, Mobile and Multimedia Networks, Crete, Greece, Jun. 2018.

\bibitem{4} 5GPPP Vision on Software Networks and 5G SN WG, Jan. 2017.

\bibitem{5} C. Liaskos, S. Nie, A. Tsioliaridou, A. Pitsillides, S. Ioannidis, and I. F. Akyildiz, “A New Wireless Communication Paradigm Through Software-Controlled Metasurfaces”, IEEE Communications Magazine, Vol. 56, No. 9, pp. 162-169, Sep. 2018.

\bibitem{6} L. Subrt and P. Pechac, “Controlling Propagation Environments Using Intelligent Walls”, European Conference on Antennas and Propagation, Prague, Czech Republic, Mar. 2012.

\bibitem{7} L. Subrt and P. Pechac, “Intelligent Walls as Autonomous Parts of Smart Indoor Environments”, IET Communications, Vol. 6, No. 8, pp. 1004-1010, May 2012.

\bibitem{8} X. Tan, Z. Sun, J. M. Jornet, and D. Pados, “Increasing Indoor Spectrum Sharing Capacity using Smart Reflect-Array”, IEEE International Conference on Communications, Kuala Lumpur, Malaysia, May 2016.

\bibitem{9} O. Abari, D. Bharadia, A. Duffield, and D. Katabi, “Enabling High-Quality Untethered Virtual Reality”, USENIX Symposium on Networked Systems Design and Implementation, Boston, USA, Mar. 2017.

\bibitem{11} A. Welkie, L. Shangguan, J. Gummeson, W. Hu, and K. Jamieson, “Programmable Radio Environments for Smart Spaces”, ACM Workshop on Hot Topics in Networks, Palo Alto, USA, Nov. 30 - Dec. 1, 2017.

\bibitem{12} R. Chandra and K. Winstein, “Programmable Radio Environments for Smart Spaces - HotNets-XVI Dialogue”, ACM Workshop on Hot Topics in Networks, Palo Alto, USA, Nov. 30 - Dec. 1, 2017.

\bibitem{13} S. Hu, F. Rusek, and O. Edfors, “Beyond Massive MIMO: The Potential of Data Transmission With Large Intelligent Surfaces”, IEEE Transactions on Signal Processing, Vol. 66, No. 10, pp. 2746-2758, May 2018.

\bibitem{10} X. Tan, Z. Sun, D. Koutsonikolas, and J. M. Jornet, “Enabling Indoor Mobile Millimeter-Wave Networks Based on Smart Reflect-Arrays”, IEEE Conference on Computer Communications, Honolulu, USA, Apr. 2018.

\bibitem{14} C. Liaskos, A. Tsioliaridou, A. Pitsillides, S. Ioannidis, and I. F. Akyildiz, “Using any Surface to Realize a New Paradigm for Wireless Communications”, Communications of the ACM, Vol. 61 No. 11, pp. 30-33, Nov. 2018.

\bibitem{15} A. Tsioliaridou, C. Liaskos, and S. Ioannidis, “Towards a Circular Economy via Intelligent Metamaterials”, IEEE International Conference on Computer-Aided Modeling Analysis and Design of Communication Links and Networks, Barcelona, Spain, Sep. 2018.

\bibitem{16} N. Yu, P. Genevet, M. A. Kats, F. Aieta, J.-P. Tetienne, F. Capasso, and Z. Gaburro, “Light Propagation with Phase Discontinuities: Generalized Laws of Reflection and Refraction”, Science, Vol. 334, No. 6054, pp. 333-337, Oct. 2011.

\bibitem{17} C. L. Holloway, E. F. Kuester, J. A. Gordon, J. O'Hara, J. Booth, and D. R. Smith, “An Overview of the Theory and Applications of Metasurfaces: The Two-Dimensional Equivalents of Metamaterials”, IEEE Antennas and Propagation Magazine, Vol. 54, No. 2, pp. 10-35, Apr. 2012.

\bibitem{18} L. Spada, “Metamaterials for Advanced Sensing Platforms”, Research Journal on Optical Photonics, Vol. 1, No. 1, Oct. 2017.

\bibitem{19} T. Nakanishi, T. Otani, Y. Tamayama, and M. Kitano, “Storage of Electromagnetic Waves in a Metamaterial That Mimics Electromagnetically Induced Transparency”, Physical Review B, Vol. 87, No. 161110, Apr. 2013.

\bibitem{20} A. Silva, F. Monticone, G. Castaldi, V. Galdi, A. Alu, and N. Engheta, “Performing Mathematical Operations with Metamaterials”, Vol. 343, No. 6167, pp. 160-163, Jan. 2014.

\bibitem{21} H2020 VISORSURF project, “A Hardware Platform for Software-Driven Functional Metasurfaces”, http://www.visorsurf.eu/.


\bibitem{Blockages} T. Bai, R. Vaze, and R. W. Heath Jr., “Analysis of Blockage Effects on Urban Cellular Networks”, IEEE Transactions on Wireless Communications, Vol. 13, No. 9, pp. 5070-5083, Sep. 2014.

\bibitem{Baccelli} J. Lee and F. Baccelli, “On the Effect of Shadowing Correlation on Wireless Network Performance”, IEEE International Conference on Computer Communications, Honolulu, Hawaii, Apr. 2018.



\bibitem{83} J. G. Andrews, F. Baccelli, and R. K. Ganti, “A Tractable Approach to Coverage and Rate in Cellular Networks”, IEEE Transactions on Communications, Vol. 59, No. 11, pp. 3122-3134, Nov. 2011.

\bibitem{MDR_1} M. Di Renzo, A. Guidotti, G. E. Corazza, “Average Rate of Downlink Heterogeneous Cellular Networks over Generalized Fading Channels – A Stochastic Geometry Approach”, IEEE Transactions on Communications, Vol. 61, No. 7, pp. 3050–3071, Jul. 2013.

%
%

\bibitem{103} M. Di Renzo, “Stochastic Geometry Modeling and Analysis of Multi-Tier Millimeter Wave Cellular Networks”, IEEE Transactions on Wireless Communications, Vol. 14, No. 9, pp. 5038-5057, Sep. 2015.

\bibitem{MDR_3} W. Lu and M. Di Renzo, “Stochastic Geometry Modeling of Cellular Networks: Analysis, Simulation and Experimental Validation”, ACM International Conference on Modeling Analysis and Simulation of Wireless and Mobile Systems, Nov. 2015.


\bibitem{105} M. Di Renzo, W. Lu, and P. Guan, “The Intensity Matching Approach: A tractable Stochastic Geometry Approximation to System-Level Analysis of Cellular Networks”, IEEE Transactions on Wireless Communications, Vol. 15, No. 9, pp. 5963-5983, Sep. 2016.

\bibitem{125} M. Di Renzo, A. Zappone, T. T. Lam, and M. Debbah, “System-Level Modeling and Optimization of the Energy Efficiency in Cellular Networks - A Stochastic Geometry Framework”, IEEE Transactions on Wireless Communications, Vol. 17, No. 4, pp. 2539-2556, Apr. 2018.

\bibitem{124} M. Di Renzo, S. Wang, and X. Xi, “Modeling and Analysis of Cellular Networks by Using Inhomogeneous Poisson Point Processes”, IEEE Transactions on Wireless Communications, Vol. 17, No. 8, pp. 5162-5182, Aug. 2018.

\bibitem{126} M. Di Renzo, T. T. Lam, A. Zappone, and M. Debbah, “A Tractable Closed-Form Expression of the Coverage Probability in Poisson Cellular Networks”, IEEE Wireless Communications Letters, to appear, 2018.

\bibitem{135} M. Di Renzo, A. Zappone, T. T. Lam, and M. Debbah, “Spectral-Energy Efficiency Pareto Front in Cellular Networks: A Stochastic Geometry Framework”, IEEE Wireless Communications Letters, to appear, 2018.




\bibitem{107} A. Narayanan, S. T. Veetil, and R. K. Ganti, “Coverage Analysis in Millimeter Wave Cellular Networks with Reflections”, IEEE Global Communications Conference, Singapore, Singapore, Dec. 2017.


\bibitem{MDR_Frechet} M. Di Renzo and W. Lu, “System-level Analysis and Optimization of Cellular Networks With Simultaneous Wireless Information and Power Transfer: Stochastic Geometry Modeling”, IEEE Transactions on Vehicular Technology, Vol. 66, No. 3, pp. 2251-2275, TBA.

\end{thebibliography}
	

	

\end{backmatter}
\end{document}